# Rotational and fine structure of open-shell molecules in nearly degenerate electronic states

Jinjun Liu

Department of Chemistry, University of Louisville, 2320 S. Brook St., Louisville, KY 40292, U.S.A.

Email: j.liu@louisville.edu

March 1, 2018

**Abstract**

An effective Hamiltonian without symmetry restriction has been developed to model the rotational and fine structure of two nearly degenerate electronic states of an open-shell molecule. In addition to the rotational Hamiltonian for an asymmetric top, this spectroscopic model includes the energy separation between the two states due to difference potential and zero-point energy difference, as well as the spin-orbit (SO), Coriolis, and electron spin-molecular rotation (SR) interactions. Hamiltonian matrices are computed using orbitally and fully symmetrized case (a) and case (b) basis sets. Intensity formulae and selection rules for rotational transitions between a pair of nearly degenerate states and a nondegenerate state have also been derived using all four basis sets. It is demonstrated using real examples of free radicals that the fine structure of a single electronic state can be simulated with either a SR tensor or a combination of SO and Coriolis constants. The related molecular constants can be determined precisely only when all interacting levels are simulated simultaneously. The present study suggests that analysis of rotational and fine structure can provide quantitative insights into vibronic interactions and related effects.

## 1. Introduction

Rotational and fine structure of open-shell molecules (free radicals) are commonly studied by microwave spectroscopy, high-resolution laser spectroscopy, and, in recent years, frequency-comb-based spectroscopy. Experimentally obtained spectra of free radicals with low symmetry or without symmetry are typically simulated using an effective Hamiltonian that consists of a rotational part and an electron spin-molecular rotation (SR) part. Different electronic states are usually treated separately. This Hamiltonian is hereafter referred to as *the isolated-states Hamiltonian*. As a result, effects of perturbing terms in the molecular Hamiltonian such as the spin-orbit (SO) and Coriolis interactions are absorbed by *effective* molecular constants including rotational and SR constants. Although feasible and economical, this method often conceals the coupling mechanisms of various angular momenta and obscures the importance of vibronic interactions. In many cases, such method also hinders direct comparison between molecular constants calculated *ab initio* and those determined in fitting the experimental spectra. With the advance of modern computers, in terms of both hardware and software, it is now possible to simulate and fit rotational and fine structure of multiple electronic or vibronic states simultaneously and take into consideration the effects of vibronic and related interactions. From an experimental point of view, different mechanisms that make contributions to the same effective molecular constants can often be separated parametrically by simulating and fitting high-resolution spectra of isotopologues simultaneously if the isotopic dependence of these mechanisms are different, which is usually the case. In order to disentangle such dependences, one relies on accurate modelling of energy level structure and transition intensities that involve different interacting states. Spectroscopic models for investigations of nearly degenerate states, especially those of open-shell molecules, are therefore strongly desired.

Investigation of rotational and fine structure can also provide valuable information on intramolecular interactions, especially vibronic interactions, which are ubiquitous in molecules. Vibronic interactions arise from the breakdown of the Born-Oppenheimer approximation, and can significantly alter the energy level structure. They are responsible for many intriguing intramolecular dynamic processes as well. For example, linear and nonlinear polyatomic molecules in orbitally degenerate electronic states are subject to Renner-Teller (RT) effect[1, 2] and Jahn-Teller (JT) effect,[3, 4] respectively. In each case, vibronic interaction removes the degeneracy, lowers the molecular symmetry, and distorts the equilibrium geometry if the interaction is sufficiently strong. By comparison, nearly degenerate electronic states are often coupled by the pseudo-Jahn-Teller (pJT) interaction.[4] All these vibronic interactions affect the rotational and fine structure perturbatively.[5] Furthermore, coupling of angular momenta can split rotational energy levels too. Such uncoupling phenomena include the Λ-type doubling as well as the *L*-uncoupling (interaction between molecular rotation and the electronic orbital angular momentum),[6] and the *l*-type doubling (interaction between molecular rotation and the vibrational angular momentum) for linear molecules.[7] For nonlinear molecules, especially nonlinear free radicals, the coupling scheme is more sophisticated due to lower symmetry and more degrees of freedom for electronic and nuclear motions.[8]

The rotational and fine structure of many nonlinear free radicals in nearly degenerate electronic or vibronic states has been investigated. They can be categorized as follows:

(i) *Binary van der Waals complexes of a diatomic free radical (e.g., NO, OH) in a $^2\Pi$ state with an inert gas atom or a closed-shell molecule.*[9-18] The ground state of the free radical has two separate SO components, $^2\Pi_{1/2}$ and $^2\Pi_{3/2}$. Interaction between the two monomers lifts the degeneracy of the $\pi_x$ and $\pi_y$ orbitals of the free radical, which leads to a barrier to free orbital motion and to a quenching of the orbital angular momentum of the unpaired electron. The resulting *difference potential* further separates the $^2\Pi_{1/2}$ and $^2\Pi_{3/2}$ SO components of the free radical, a mechanism analogous to energy level splitting in linear polyatomic molecules subject to the RT effect. Note that, although unmentioned in previous works, difference in the zero-point energy ($\Delta ZPE$) between the nearly degenerate states of the dimer, especially that due to the monomer-monomer stretching mode, also contributes to the separation of these two states.

(ii) *Asymmetrically deuterated JT molecules.* Upon asymmetric deuteration, the electronic degeneracy of JT molecules is retained. However, the vibrational degeneracy and hence the vibronic degeneracy are removed due to uneven distribution of mass. $\Delta ZPE$ between the two orbitally degenerate electronic states splits the vibrational ground levels.[19] Rotational structures of asymmetrically deuterated cyclopentadienyl ($C_5H_4D$ and $C_5HD_4$),[20] the methane cation ($CH_3D^+$, $CH_2D_2^+$, $CHD_3^+$),[21, 22] and methoxy ($CH_2DO$ and $CHD_2O$)[23, 24] have been investigated.

(iii) *Asymmetrical alkyl substitution of JT molecules.* Orbital degeneracy can be removed by chemical substitution that converts JT molecules to pJT molecules. In this case, both the difference potential and $\Delta ZPE$ contribute to the energy separation of the two nearly degenerate electronic states. For example, Liu et al. investigated the $\tilde{X}$ and $\tilde{A}$-state rotational and fine structure of two free radicals in this category: isopropoxy[25, 26] and cyclohexoxy,[27, 28] both of which can be regarded as alkyl substitution of the methoxy radical.

Free radicals in all three categories that have been studied belong to $C_{2v}$ or $C_s$ point groups. Their molecular symmetry can be further lowered by asymmetric isotopic or chemical substitution, which leads to more complications. The objective of the present work is therefore to derive, without any symmetry restriction, an effective Hamiltonian and its matrix elements for modeling the rotational and fine structure of an open-shell molecule in nearly degenerate electronic states. Furthermore, we will derive the intensity formulae for rotational transitions that involve one pair of such states and determine the selection rules. Although in this paper we limit ourselves to the case of double orbital (near-)degeneracy with $\Lambda = \pm 1$ as well as double spin multiplicity ($S = 1/2$ and $\Sigma == \pm 1/2$), the method presented here is applicable to the general case of multiple spin-vibronic states, which will be briefly discussed later.

## 2. Choice of coordinate systems and basis sets

Angular momenta considered in the present work include:

$S$, the total spin angular momentum of the electrons,

$L$, the total orbital angular momentum of the electrons,

$R$, the rotational angular momentum of the nuclei, and

$G$, the vibrational angular momentum of the nuclei, which is associated with two (nearly) degenerate vibrational modes that couple the two electronic states, i.e., degenerate RT-active bending modes for linear polyatomic molecules, and degenerate JT-active or nearly degenerate pJT-active modes for nonlinear molecules.

The total angular momentum excluding the total nuclear spin is $\boldsymbol{J} = \boldsymbol{S} + \boldsymbol{L} + \boldsymbol{R} + \boldsymbol{G}$. Using the Hund's case (b) coupling scheme, $\boldsymbol{J} = \boldsymbol{S} + \boldsymbol{N}$, where $\boldsymbol{N} = \boldsymbol{L} + \boldsymbol{R} + \boldsymbol{G}$ is the total angular momentum excluding spin.

In constructing the basis sets and the Hamiltonian matrices, the following assumptions are made for convenience:

1) Although in the most general case the molecule under discussion belongs to the chiral nonsymmetric $C_1$ point group, the orbital and spin wave functions of the unpaired electron are localized and both are of (near) double degeneracy. They can be approximated by the one-electron wave functions of a diatomic molecule in a $^2\Pi_{1/2,3/2}$ state or their linear combinations.

2) Similarly, the wave functions of the two (nearly) degenerate vibrational modes responsible for $\boldsymbol{G}$ can be approximated by the RT-active bending modes of a linear polyatomic molecule or their linear combinations. If a $C_s$ plane is present, these two modes are within and out of the symmetry plane, respectively.

3) Although $\boldsymbol{L}$ and $\boldsymbol{G}$ are quenched due to removal of the cylindrical symmetry, the force field and mass are so distributed that there still exists a plane, with respect to which both the orbital and the vibrational wave functions retain their symmetry ($A'$ or $A''$). $\boldsymbol{L}$ and $\boldsymbol{G}$ are therefore contained in this symmetry plane.

4) It is further assumed that $\boldsymbol{L}$ and $\boldsymbol{G}$ have the same orientation.

As will be demonstrated later, the methods developed in the present work can be extended to the cases of higher degrees of orbital, spin, and/or vibrational degeneracy (see Section 8). Moreover, simple modifications to the effective Hamiltonian are sufficient if $L$ and $G$ have different orientations (see Section 7). The assumptions listed above are therefore mainly for the sake of convenience of discussion.

Given the forgoing assumptions, one can construct an internal axis system (IAS) ($x', y', z'$) in which the $z'$ axis coincides with $L$ and $G$ (see Figure 1a). The $x'$ axis is chosen to lie within the foregoing symmetry plane (see Assumption 3), and the $y'$ axis is perpendicular to it. Projections of $J$, $S$, $N$, $L$, $G$ onto the $z'$ axis are denoted by $P$, $\Sigma$, $K$, $\Lambda$, $l$, respectively. The nonrotating-molecule basis set in the Hund's case (a) coupling scheme is:[29]

$$|L,\Lambda,v,l,S,\Sigma\rangle = f_{L|\Lambda|}\exp\{i\Lambda\varphi_e\}\rho_{v|l|}\exp\{il\varphi_n\}\begin{cases}\alpha \\ \beta\end{cases}, \tag{1}$$

where $\varphi_e$ is the azimuthal angle of the unpaired electron about the $z'$ axis with respect to the $z'x'$ plane, while $\varphi_n$ is the azimuthal angle of the atom on which the unpaired electron is localized (see Figure 1b). $\Lambda = \pm 1$ represent the situations where $L$ rotates counterclockwise or clockwise around the $z'$ axis and the electronic orbital basis functions are $\exp\{\pm i\varphi_e\}$ for $\Lambda = \pm 1$, respectively. Dependence of the orbital basis functions on other coordinates is contained in $f_{L|\Lambda|}$. In the vibrational basis functions, $\rho_{v|l|}$ is the amplitude of the two-dimensional harmonic oscillator with which $G$ is associated. Subscript $v$ is the vibrational quantum number. In the present work, we limit ourselves to the case of vibrational ground levels so that $v = 0$, and $l = 0$. $\alpha$ and $\beta$ are the electron spin basis functions denoting the situations where the projection of $S$ is along the $z'$ and $-z'$ axes ($\Sigma == \pm 1/2$), respectively.

As will be shown later, it is convenient to derive the SO and Coriolis Hamiltonians in IAS, but the rotational and SR ones are cumbersome using IAS.[23] We therefore will derive the SO and Coriolis Hamiltonians first in IAS, then convert IAS to the principal axis system (PAS) ($a,b,c$). Rotational and SR Hamiltonians will be derived in PAS directly. The relation between IAS and PAS is illustrated in Figure 1a. In the present work, the $I^r$ representation is adopted so that the $z$, $x$, $y$ axes correspond to the $a$, $b$, $c$ axes, respectively. The polar angle and the azimuthal angle of the $z'$ axis (i.e., the direction of $L$ and $G$) in PAS are denoted with $\theta$ and $\phi$, respectively (see Figure 1a). $z'$ axis in IAS can be converted to $z$ axis in PAS using a unitary transformation matrix

$$U = \begin{pmatrix} U_{zz'} & U_{zx'} & U_{zy'} \\ U_{xz'} & U_{xx'} & U_{xy'} \\ U_{yz'} & U_{yx'} & U_{yz'} \end{pmatrix} = \begin{pmatrix} \cos\theta & -\sin\theta & 0 \\ \sin\theta\cos\phi & \cos\theta\cos\phi & -\sin\phi \\ \sin\theta\sin\phi & \cos\theta\sin\phi & \cos\phi \end{pmatrix}.$$

Let's first inspect the symmetry of the nonrotating molecule basis function. In the present paper, we consider only the electronic basis function derived from the one-electron atomic configuration $np$, viz., $|\Lambda=\pm 1\rangle$, which transforms as follows under the operation $\hat{\sigma}_v(x'z')$, i.e., reflection with respect to the $z'x'$ plane:[30]

$$\hat{\sigma}_v(x'z')|\Lambda\rangle = |-\Lambda\rangle. \qquad (2)$$

A symmetrized basis set $|\Gamma=\pm 1\rangle$ can therefore be constructed from linear combinations of $|\pm\Lambda\rangle$:

$$|\Gamma=\pm 1\rangle = \tfrac{1}{\sqrt{2}}\left(|\Lambda=+1\rangle \pm |\Lambda=-1\rangle\right). \qquad (3)$$

$|\Gamma = \pm 1\rangle$ transform as follows under $\hat{\sigma}_v(x'z')$:

$$\hat{\sigma}_v(x'z')|\Gamma\rangle = \Gamma|\Gamma\rangle. \tag{4}$$

Eigenfunctions $|\Gamma = \pm 1\rangle$ belong to the $A'$ and $A''$ irreducible representations of the $C_s$ symmetry group, respectively. $\Gamma = \pm 1$ therefore denotes the reflection symmetry of the orbital wave function.

The rotating-molecule basis set can be constructed as direct product of $|\Lambda\rangle$ and Hund's case (a) or (b) spin-rotational basis sets. If the SO interaction is strong and the SO splitting dominates the energy separation between the two electronic states, the Hund's case (a) basis set $|\Lambda, J, P, S, \Sigma\rangle$ is suitable. If the SO interaction is relatively weak and the energy separation between the two electronic states is mainly due to the difference potential and/or $\Delta ZPE$, a Hund's case (b) basis set $|\Lambda, J, N, K, S\rangle$ is easier to apply. To simplify the Hamiltonian matrices, a more convenient approach is to combine the spin-rotational basis sets with the symmetrized orbital basis set $|\Gamma\rangle$ in analogy to open-shell diatomic molecules. The orbitally symmetrized spin-ro-orbital basis set is:

$$|\Gamma, J, P, S, \Sigma\rangle = \tfrac{1}{\sqrt{2}}\left[|\Lambda = +1, J, P, S, \Sigma\rangle + \Gamma|\Lambda = -1, J, P, S, \Sigma\rangle\right] \tag{5}$$

using the case (a) spin-rotational basis set, and

$$|\Gamma, J, N, K, S\rangle = \tfrac{1}{\sqrt{2}}\left[|\Lambda = +1, J, N, K, S\rangle + \Gamma|\Lambda = -1, J, N, K, S\rangle\right] \tag{6}$$

using the case (b) spin-rotational basis set. Eqs. (5) and (6) will be referred to as *orbitally symmetrized case (a) and (b) basis sets*, respectively.

The two spin-ro-orbital basis sets above are *not* eigenfunctions of $\hat{\sigma}_v(x'z')$. In a representation with these orbitally symmetrized basis functions, all eigenfunctions are linear combination of opposite $P$ and $S$ (in the case (a) basis set), or opposite $K$ (in the case (b) basis set). (See Section 5.1 for details.) One may combine basis functions with opposite $P$ and $S$, or $K$, to construct spin-ro-orbital basis sets that belong to the irreducible representations of the $C_s$ group. For this purpose, symmetry properties of the spin-rotational basis function are revisited as follows.

The effect of $\hat{\sigma}_v(x'z')$ on the case (a) spin-rotational basis functions is:[30, 31]

$$\hat{\sigma}_v(x'z')|J,P\rangle = (-1)^{J-P}|J,-P\rangle$$
$$\hat{\sigma}_v(x'z')|S,\Sigma\rangle = (-1)^{S-\Sigma}|S,-\Sigma\rangle \quad . \tag{7}$$

The case (b) spin-rotational basis function transforms as:

$$\hat{\sigma}_v(x'z')|J,N,K,S\rangle = (-1)^{N-K}|J,N,-K,S\rangle \tag{8}$$

under $\hat{\sigma}_v(x'z')$.

We introduce a new symmetry symbol $s$ that denotes the symmetry of the spin-rotational wave function with respect to $\hat{\sigma}_v(x'z')$. For the case (a) basis set, $s = (-1)^{J-P+S-\Sigma}$, while for the case (b) basis set, $s = (-1)^{N-K}$.

Given transformation properties of the basis sets (Eqs. 4, 7, 8), fully symmetrized basis sets can be constructed by introducing the reflection symmetry of the overall spin-ro-orbital basis function with respect to the $x'z'$ plane, denoted with $\wp = \Gamma s = \pm 1$:

(a) *Fully symmetrized case (a) basis set*:[32]

$$\left|J,\bar{P},S,\bar{\Sigma},\wp\right\rangle = \tfrac{1}{\sqrt{2}}\left[\left|\Lambda=+1,J,P,S,\Sigma\right\rangle + \wp(-1)^{J-P+S-\Sigma}\left|\Lambda=-1,J,-P,S,-\Sigma\right\rangle\right]. \tag{9}$$

In the present work, a bar is used to indicate a quantum number in a fully symmetrized basis set that is a mixture of two opposite values of a quantum number in the corresponding spin-rotational basis sets.

(b) *Fully symmetrized case (b) basis set:*

$$\left|J,N,\bar{K},S,\wp\right\rangle = \tfrac{1}{\sqrt{2}}\left[\left|\Lambda=+1,J,N,K,S\right\rangle + \wp(-1)^{N-K}\left|\Lambda=-1,J,N,-K,S\right\rangle\right]. \tag{10}$$

The two basis functions in Eqs. (9) and (10) are eigenfunctions of $\hat{\sigma}_v(x'z')$ and transform as follows:

$$\hat{\sigma}_v(x'z')\left|J,\bar{P},S,\bar{\Sigma},\wp\right\rangle = \wp\left|J,\bar{P},S,\bar{\Sigma},\wp\right\rangle, \tag{11}$$

$$\hat{\sigma}_v(x'z')\left|J,N,\bar{K},S,\wp\right\rangle = \wp\left|J,N,\bar{K},S,\wp\right\rangle. \tag{12}$$

For linear and planar polyatomic molecules, the net effect of the symmetry operation $\hat{\sigma}_v(x'z')$ acting on a complete basis set function is equivalent to that of the laboratory-fixed inversion operation (denoted with $\hat{E}^*$ or $\hat{I}$ by different authors).[30, 33, 34] Therefore, $\wp$ is the *parity* of rotational energy levels in these special cases. For most nonplanar molecules $\hat{E}^*$ is unfeasible. Following Longuet-Higgins' rule that the molecular symmetry group is composed of feasible elements only,[35] the parity for these molecules is not a useful quantum number anymore, and $\wp$ merely indicates the reflection symmetry or "eigenvalues" of the $\hat{\sigma}_v(x'z')$ operator.

The spin-rotational case (a) and (b) basis sets are related to each other by a unitary transformation.[8, 36, 37] Obviously, combination with the orbital basis functions as in the orbitally symmetrized basis set (Eqs. 5 and 6) doesn't change this relation. One therefore has:

$$|\Gamma, J, N = J \pm \tfrac{1}{2}, K, S\rangle = \left(\frac{J \mp K + \tfrac{1}{2}}{2J+1}\right)^{1/2} |\Gamma, J, P = K + \tfrac{1}{2}, S, \Sigma = +\tfrac{1}{2}\rangle \mp \left(\frac{J \pm K + \tfrac{1}{2}}{2J+1}\right)^{1/2} |\Gamma, J, P = K - \tfrac{1}{2}, S, \Sigma = -\tfrac{1}{2}\rangle. \qquad (13)$$

Combining Eq. (13) with the unitary transformations (i.e., linear combinations) in Eqs. (9) and (10), one obtains the relation between the fully symmetrized case (a) and (b) basis sets:

$$|J, N = J \pm \tfrac{1}{2}, \overline{K}, S, \wp\rangle = \left(\frac{J \mp \overline{K} + \tfrac{1}{2}}{2J+1}\right)^{1/2} |\Gamma, J, \overline{P} = \overline{K} + \tfrac{1}{2}, S, \overline{\Sigma} = +\tfrac{1}{2}\rangle \mp \left(\frac{J \pm \overline{K} + \tfrac{1}{2}}{2J+1}\right)^{1/2} |\Gamma, J, \overline{P} = \overline{K} - \tfrac{1}{2}, S, \overline{\Sigma} = -\tfrac{1}{2}\rangle, \qquad (14)$$

which has the same form as for the orbitally symmetrized ones (Eq. 13) except that $\Gamma$ is replaced with $\wp$.

As will be shown below, it is more convenient to derive the matrices of the SO and the Coriolis Hamiltonians in the case (a) basis sets, which can be converted to the case (b) basis sets using the unitary transformations above.

### 3. Effective Hamiltonian

The effective Hamiltonian proposed in the present work, hereafter referred to as *the coupled-states Hamiltonian*, consists of five terms.[31, 32, 38]

$$H_{eff} = H_q + H_{SO} + H_C + H_r + H_{SR}. \tag{15}$$

The first term $H_q$ is associated with the energy separation between the rotationless levels. $H_{SO}$, $H_C$, $H_r$, and $H_{SR}$ are the SO, Coriolis, rotational, and SR terms, respectively.

(i) $H_q$: This term of the effective Hamiltonian has two origins: quenching of the electronic orbital angular momentum $L$ that causes the difference potential, and the difference between the zero-point energies of the two electronic states ($\Delta ZPE$). We shall discuss the former mechanism first.

For a nonrotating molecule with cylindrical symmetry, an energy degeneracy exists between molecular orbitals that are perpendicular to the symmetry axis, e.g., the $\pi_x$ and $\pi_y$ orbitals of a diatomic molecule or a linear polyatomic molecule. Another example is the $e_x$ and $e_y$ orbitals of a $C_{nv}$ ($n \geq 3$) molecule. For molecules with lower symmetry, the orbital degeneracy is lifted and $L$ is quenched by Coulombic interactions. Such interactions lead to a barrier to free orbital motion. Analogous to the RT effect,[1, 29] the quenching potential can be expanded as a Fourier series in the azimuthal coordinate about the $z'$ axis:[9]

$$H_q = \sum_n \varepsilon_n \cos(n\Delta\varphi) = \varepsilon_0 + \varepsilon_1 \cos(\Delta\varphi) + \varepsilon_2 \cos(2\Delta\varphi) + ..., \tag{16}$$

where $\Delta\varphi = \varphi_e - \varphi_n$ (see Figure 1b). The $n^{th}$-order term of $H_q$ connects basis functions with $\Delta\Lambda = n$. Specifically, the zeroth-order term shifts all states by the same magnitude and hence is absorbed by the term value. The first-order term connects states with $\Delta\Lambda = \pm 1$. Its effect on the rotational structure can be neglected.[9] The second-order term connects the $|\Lambda = \pm 1\rangle$ states. Contribution from higher-order terms are neglected. Overall, the matrix form of $H_q$ in the representation of $|\Lambda = \pm 1\rangle$ is:

$$H_q = \frac{\varepsilon_2}{2}\begin{pmatrix} 0 & 1 \\ 1 & 0 \end{pmatrix}. \tag{17}$$

in which normalization conditions are applied. Matrix form of $H_q$ in the representation of $|\Gamma\rangle$ can be derived using a unitary matrix $S = \frac{1}{\sqrt{2}}\begin{pmatrix} 1 & 1 \\ 1 & -1 \end{pmatrix}$ that transforms the $|\Lambda = \pm 1\rangle$ basis set to the $|\Gamma = \pm 1\rangle$ basis set (see Eq. 3). The result is:

$$H_q = \frac{\varepsilon_2}{2}\begin{pmatrix} 1 & 0 \\ 0 & -1 \end{pmatrix}. \tag{18}$$

It is therefore evident that $\varepsilon_2$ is the *difference potential* between the $|\Gamma = +1\rangle$ ($A'$) and $|\Gamma = -1\rangle$ ($A''$) basis functions. A positive $\varepsilon_2$ would imply that the $A'$ state is above the $A''$ state, and vice versa.

When the vibrational motion is taken into consideration, $\Delta ZPE$ further separates the $A'$ and the $A''$ diabatic states. The two mechanisms that contribute to $H_q$ – different potential and $\Delta ZPE$ - have the same form and cannot be determined independently by analyzing the rotational and fine structure. They can be separated parametrically by the use of other information such as (i) isotopic variation of the overall energy separation, and (ii) vibronic analysis. We define a new molecular constant $\Delta E_0 = \varepsilon_2 + \Delta ZPE$ to account for the combined effect of these two mechanisms and $\Delta E_0$ represents the overall energy separation between the two rotationless levels in the absence of SO and Coriolis interactions. With the new definition of $H_q$, $\varepsilon_2$ in Eqs. (17) and (18) needs to be replaced with $\Delta E_0$.

Hougen[9, 10, 31] introduced two ladder operators $\mathcal{L}_\pm^2$, which are in essence normalized $(L_{x'} \pm iL_{y'})^{2\Lambda}$. For $|\Lambda = \pm 1\rangle$ states, $\mathcal{L}_\pm^2 = |\Lambda = \pm 1\rangle\langle\Lambda = \mp 1|$. $H_q$ can therefore be conveniently written as:

$$H_q = \tfrac{1}{2}\Delta E_0 \left(\mathcal{L}_+^2 + \mathcal{L}_-^2\right). \tag{19}$$

(ii) $H_{SO}$: With the aforementioned assumptions, the SO Hamiltonian can be written as:

$$H_{SO} = \sum_\alpha a_\alpha L_\alpha S_\alpha, \tag{20}$$

where $\alpha$ is a Cartesian coordinate, and $a$'s are SO constants. In IAS, $H_{SO}$ may take the form:

$$H_{SO} = a_\| L_z S_{z'} + \tfrac{1}{2}a_\perp^+(L_x S_{x'} + L_y S_{y'}) + \tfrac{1}{2}a_\perp^-(L_x S_{x'} - L_y S_{y'}). \tag{21}$$

The effect of $(L_x S_{x'} \pm L_y S_{y'})$ is negligible to the first-order of approximation.[39] Their second-order contribution is absorbed by the SR constants.[31] Using the unitary transformation matrix $U$ that converts IAS to PAS, it is not difficult to show that in PAS:[23, 40]

$$H_{SO} = a_\| L_{z'}(\cos\theta S_z + \sin\theta\cos\phi S_x + \sin\theta\sin\phi S_y), \tag{22}$$

where $\theta$ and $\phi$ are the polar angle and the azimuthal angle of $\boldsymbol{L}$ as well as $\boldsymbol{G}$ in PAS, respectively. (See Figure 1a.)

In spectroscopic analysis, it is often advantageous to partition the *effective* SO constant into different factors (see Section 7). For nonrotating diatomic molecules, the orbital angular momentum projection quantum number $\Lambda$ is a good quantum number in either the case (a) or case (b) coupling scheme. The expectation value of the orbital angular momentum $L_{z'}$ in the orbital basis function $|L, \Lambda\rangle$ is equal to $\Lambda$, i.e., $\langle L, \Lambda | L_{z'} | L, \Lambda \rangle = \Lambda$. In RT, JT and pJT-active states, $\Lambda$ ceases to be a good quantum number, and the SO interaction is quenched not only electronically[41,42] but also vibronically[3,19-20] through mixing of vibronic basis functions. For instance, the ground-state SO splitting of the OH($\tilde{X}^2\Pi$) radical[43] is -139 cm$^{-1}$, whereas it is reduced to -61.5 cm$^{-1}$ for the CH$_3$O($\tilde{X}^2E$) radical[44] upon methyl substitution. Here we adopt the notation $a\zeta_e d$ for the *effective* (quenched) SO constant of vibronically coupled states, where $a$ is the atomic-like SO constant, approximated by that of a diatomic reference molecule. (The "∥" subscript is omitted to follow convention.) $\zeta_e$ is the electronic quenching factor, and $d$ is the vibronic quenching factor, a.k.a., the Ham reduction factor.[45] Overall, $\zeta_e d$ is the expectation value of $L_{z'}$ in the vibronic eigenfunction. The SO Hamiltonian (Eq. 22) thus reduces to:

$$H_{SO} = a\zeta_e d \mathscr{L}_{z'}(\cos\theta S_z + \sin\theta \cos\phi S_x + \sin\theta \sin\phi S_y) \tag{23}$$

where $\mathscr{L}_{z'} = |\Lambda = +1\rangle\langle\Lambda = +1| - |\Lambda = -1\rangle\langle\Lambda = -1|$ can be regarded as the normalized $L_{z'}$ operator. Action rules of Hougen's operators on different basis functions are summarized in the Supplementary Materials (see Section S.1).

(iii) $H_C + H_r$: Following Watson[46] but adopting different symbols for angular momenta the kinetic ro-vibrational Hamiltonian of a nonlinear molecule can be written as:

$$H_{KE} = \tfrac{1}{2}\sum_{\alpha,\beta}(J_\alpha - S_\alpha - L_\alpha - G_\alpha)\mu_{\alpha\beta}(J_\beta - S_\beta - L_\beta - G_\beta)$$
$$= \tfrac{1}{2}\sum_{\alpha,\beta}[N_\alpha - (L_\alpha + G_\alpha)]\mu_{\alpha\beta}[N_\beta - (L_\beta + G_\beta)] \quad , \tag{24}$$

where $\alpha$ and $\beta$ are Cartesian coordinates, and $\boldsymbol{\mu}$ is an effective reciprocal inertia tensor, viz., $\mu_{\alpha\beta} = I^{-1}_{\alpha\beta}$, where $\boldsymbol{I}$ is the moment of inertia tensor.

In Eq. (24) the $(L_\alpha + G_\alpha)(L_\beta + G_\beta)$ terms shift all rotational levels by the same magnitude and are absorbed by the term value. The $-N_\alpha \mu_{\alpha\beta}(L_\beta + G_\beta)$ cross terms contribute to the Coriolis interaction. In IAS,

$$H_C = -\sum_{\alpha,\beta=x',y',z'} \mu_{\alpha\beta} N_\alpha (L_\beta + G_\beta) \tag{25}$$

Similar to the SO Hamiltonian, contribution from $L_{x'} + G_{x'}$ and $L_{y'} + G_{y'}$ can be neglected. Transformed into PAS, $H_C$ takes the form of:

$$H_C = -2(\cos\theta B_z N_z + \sin\theta\cos\phi B_x N_x + \sin\theta\sin\phi N_y S_y)(L_{z'} + G_{z'}), \tag{26}$$

where $B_\alpha$ is the rotational constant about the $\alpha$ principal axis, viz., $A \geq B \geq C$ following convention. We follow the convention for JT molecules and denote with $\zeta_t$ the expectation value of the sum of the orbital and vibrational angular momenta $(L_{z'} + G_{z'})$ in the vibronic eigenfunction. The Coriolis Hamiltonian (Eq. 26) reduces to:

$$H_C = -2\zeta_t \mathcal{L}_{z'}(\cos\theta B_z N_z + \sin\theta\cos\phi B_x N_x + \sin\theta\sin\phi B_y N_y). \tag{27}$$

The $\frac{1}{2}N_\alpha \mu_{\alpha\beta} N_\beta$ terms constitute the rotational Hamiltonian and is diagonal with respect to $\Lambda$ and $\Sigma$. In PAS $H_r$ can be simplified to the well-known form of:

$$H_r = \sum_{\alpha=x,y,z} B_\alpha N_\alpha^2. \tag{28}$$

Note that in case (a) basis sets, $N_\alpha = J_\alpha - S_\alpha$. The rotational Hamiltonian therefore contains terms that are proportional to $J_\alpha^2$, $S_\alpha^2$, and $-2J_\alpha S_\alpha$ (the spin-uncoupling terms), respectively.

(iv) $H_{SR}$: Van Vleck[47] first derived the form of the effective SR Hamiltonian from a consideration of the mixing of electronic states by the combined effects of the SO and Coriolis terms in the rotational Hamiltonian. $H_{SR}$ is diagonal with respect to $\Lambda$ and takes the form:

$$H_{SR} = \frac{1}{2}\sum_{\alpha,\beta} \varepsilon_{\alpha\beta}(N_\alpha S_\beta + S_\beta N_\alpha), \tag{29}$$

where $\varepsilon_{\alpha\beta}$ are SR constants. Note that $H_{SR}$ contains both $\frac{1}{2}J_\alpha S_\beta$ and $\frac{1}{2}S_\alpha S_\beta$ terms in case (a) basis sets. Brown and Sears[48] have demonstrated that for $C_1$ molecules, there are six determinable quadratic SR constants ($\varepsilon_{zz}$, $\varepsilon_{xx}$, $\varepsilon_{yy}$, $|(\varepsilon_{zx} + \varepsilon_{xz})/2|$, $|(\varepsilon_{xy} + \varepsilon_{yx})/2|$, and $|(\varepsilon_{yz} + \varepsilon_{zy})/2|$), which will be used in the present work. For the latter three, only absolute values can be determined from spectral simulation and fitting.

## 4. Hamiltonian matrix elements

In this section, we will derive the matrix elements of all Hamiltonian terms in Eq. (15), starting with the vibronic quenching term $H_q$. Since neither rotational nor spin angular momenta are involved in $H_q$, its matrix is diagonal with respect to all angular momentum quantum numbers except $\Lambda$. Regardless of the spin-rotational basis set, matrix elements of $H_q$ in the orbtially symmetrized basis sets (Eqs. 5 and 6) can be easily obtained using Eq. (18) and written as:

$$\left\langle \Gamma' \middle| H_q \middle| \Gamma \right\rangle = \tfrac{1}{2} \delta_{\Gamma,\Gamma'} \Gamma \Delta E_0, \tag{30}$$

where $\delta_{\Gamma,\Gamma'} = 1$ if $\Gamma' = \Gamma$, and $\delta_{\Gamma,\Gamma'} = 0$ if $\Gamma' \neq \Gamma$. In the orbitally symmetrized basis sets, $H_q$ doesn't affect the rotational and fine structure of the $A'$ and the $A''$ eigenstates. If $\Delta E_0 > 0$, $H_q$ shifts all rotational levels in the $A'$ state ($\Gamma = +1$) upward and the $A''$ state ($\Gamma = -1$) downward by $|\Delta E_0|/2$, and vice versa.

In the fully symmetrized basis sets, the reflection symmetry of the spin-rotational wave function, $s$, needs to be taken into account. $\Gamma$ in Eq. (30) needs to be replaced with $s\wp$. Therefore, the nonzero matrix elements of $H_q$ are:

$$\left\langle J, -\bar{P}, S, -\bar{\Sigma}, \wp \middle| H_q \middle| J, \bar{P}, S, \bar{\Sigma}, \wp \right\rangle = \tfrac{1}{2} \wp (-1)^{J-\bar{P}+S-\bar{\Sigma}} \Delta E_0 \tag{31}$$

in the fully symmetrized case (a) basis, and

$$\left\langle J, N, -\bar{K}, S, \wp \middle| H_q \middle| J, N, \bar{K}, S, \wp \right\rangle = \tfrac{1}{2} \wp (-1)^{N-\bar{K}} \Delta E_0 \tag{32}$$

in the fully symmetrized case (b) basis.

Matrix elements of other Hamiltonian terms in the orbitally or fully symmetrized case (a) basis sets can be computed by first deriving action rules of angular momentum operators $\boldsymbol{J}$ and $\boldsymbol{S}$, as well as $\mathcal{L}_z$, on the basis functions. This can be done by employing the common angular momentum action rules on the spin-rotational case (a) basis set[49] and applying them to Eqs. (5) and (9), respectively.[37, 50] The action rules are listed in the Supplementary Materials (see S.2 and S.3). Several observations are made:

1) $\mathcal{L}_z$ connects $\Gamma$ states to $-\Gamma$ states (in the orbitally symmetrized basis set), and $\wp$ states to $-\wp$ states (in the fully symmetrized basis set). It doesn't affect quantum numbers of angular momenta ($J$ and $S$) and their projections ($P$ and $\Sigma$).

2) In the orbitally symmetrized basis set, both $J_\alpha$ and $S_\alpha$ ($\alpha = x, y, z$) connect $\Gamma$ states to $\Gamma$ states.

3) In the fully symmetrized basis set, $J_z$, $J_x$ and $S_z$ $S_x$ connect $\wp$ states to $-\wp$ states, whereas $J_y$ and $S_y$ do not switch $\wp$.

4) If $\Gamma$ and $\wp$ are disregarded, action rules of $J_\alpha$ and $S_\alpha$ on both the orbitally and the fully symmetrized basis sets have the same form as those on the spin-rotational basis set ($|J, P, S, \Sigma\rangle$).

Using the products of these action rules, it is not difficult to compute the matrix elements of the effective Hamiltonian in both the orbitally and the fully symmetrized case (a) basis sets, which are listed in the Supplementary Materials (see Section S.4 and S.5, respectively).

To compute the Hamiltonian matrices in the symmetrized case (b) basis sets, one may follow the same strategy and derive first the action rules of $\mathscr{L}_z$, $N$ and $S$ on the case (b) basis functions. Those of $N$ are more straightforward than $S$. Nevertheless, action rules of these angular momentum operators and their products on the spin-rotational case (b) basis set can be found in Refs. [51-53]. Using the unitary transformation in Eqs. (6) and (10), action rules on the symmetrized case (b) basis sets can be derived.

The approach outlined above for computing the Hamiltonian matrices in case (b) basis sets, although elementary, is tedious for $H_{SO}$ and $H_{SR}$ because of the presence of the spin angular momentum operator. Hamiltonian elements can be computed more conveniently as follows.

(i) One can convert the matrix of the SO Hamiltonian in the orbitally and fully symmetrized case (a) basis sets (see Section S.4 and S.5) to the corresponding symmetrized case (b) basis sets using Eqs. (13) and (14) and the inverse transforms. This method applies to the Coriolis Hamiltonian as well, though direct computation using the action rules of the angular momentum operators is more convenient.

(ii) Matrix elements of the SR as well as rotational Hamiltonians in the spin-rotational case (b) basis set have been computed previously[47, 54, 55] (see, for example, Table II of Ref. [47] and Table I of Ref. [55]). Using the unitary transformations in Eqs. (6) and (10), it can be proven that each matrix element of $H_r$ and $H_{SR}$ in the symmetrized case (b) basis sets is identical to that in the case (b) spin-rotational basis set if the symmetry quantum numbers $\Gamma$ and $\wp$ are disregarded. These two Hamiltonians are diagonal with respect to $\Gamma$ in the orbitally symmetrized basis set because the orbital angular momentum $\boldsymbol{L}$ is not involved. In the fully symmetrized basis set, the rotational Hamiltonian is diagonal with respect to $\wp$. So is $H_{SR}$ except for those terms that are proportional to $\frac{1}{2}(N_x S_y + N_y S_x)$ or $\frac{1}{2}(N_y S_z + N_z S_y)$, which connect $\wp = \pm 1$ basis functions. This can be understood based on Observations (2) and (3) stated above. (Note that $N_\alpha = J_\alpha - S_\alpha$.) These terms are purely imaginary due to the action rules of $S_y$ (see Sections S.2 and S.3).

All Hamiltonian elements in the orbitally and the fully symmetrized case (b) basis sets are listed in the Supplementary Materials (see Section S.6 and S.7, respectively).

## 5. Discussion of the effective Hamiltonian and the energy level structure

### 5.1. Structure of Hamiltonian matrices and symmetry of energy levels

Figure 2 illustrates the matrix structure of the effective Hamiltonian (Eq. 15) in all four representations. $H_q$ is diagonal with respect to $\Gamma$ (in orbitally symmetrized basis sets) or $\wp$ (in fully symmetrized basis sets). Moreover, it has only diagonal elements in orbitally symmetrized basis sets. In the fully symmetrized case (a) basis set, it connects basis functions with opposite $\bar{P}$ and $\bar{\Sigma}$, while it connects basis functions with $\pm\bar{K}$ in the fully symmetrized case (b) basis set. SO and Coriolis terms are off-diagonal with respect to $\Gamma$. The $z$- and $x$-components of these two

Hamiltonian terms are diagonal with respect to $\wp$, while their $y$-components connect basis functions with $\pm\wp$.

For weakly coupled electronic states with $a\zeta_e d \ll \Delta E_0$, the orbitally symmetrized basis set is a better representation since $H_q$ is automatically diagonalized. In the limiting case of zero SO and Coriolis interactions, orbital wave functions of the two electronic states are unmixed and have $A'$ and $A''$ character for the $\Gamma = +1$ and $\Gamma = -1$ states, respectively. In this limiting case, eigenfunctions of all rotational levels are mixtures of equally weighted $\pm P$ and $\pm \Sigma$ basis functions in the totally symmetrized case (a) basis set, and mixtures of equally weighted $\pm K$ basis functions in the totally symmetrized case (b) basis set. The "phase" of the mixing is determined by $s\Gamma$ or $s\wp$ in each case.

If the molecule belongs to the $C_s$ point group or other higher-symmetry point groups that have reflection symmetry, the $y$-components of SO and Coriolis terms vanish, and so do the SR constants $(\varepsilon_{xy} + \varepsilon_{yx})/2$ and $(\varepsilon_{yz} + \varepsilon_{zy})/2$ for the $(N_x S_y + N_y S_x)$ and $(N_y S_z + N_z S_y)$ terms, respectively, all of which connect the $\wp = \pm 1$ basis functions. $H_{eff}$ is therefore diagonal with respect to $\wp$. Fully symmetrized basis sets are therefore more convenient for computing the matrix elements and the matrix diagonalization.

It's worth pointing out that the JT- (or pJT-)induced ro-vibronic term[5, 31, 32, 56] $H_{JT}$ is not included *explicitly* in the present work. This term has two contributions.[56] The "genuine" (p)JT-contribution to $H_{JT}$ is a (p)JT-induced geometric distortion of the equilibrium structure and the resulting difference in the moment of inertia tensor between the two (nearly) degenerate states. The other contribution is similar to $L$-uncoupling and is caused by Coriolis interaction between the (p)JT states and other electronic states. Although Hamiltonian matrix elements due to these two mechanisms have identical forms, their contributions to the corresponding molecular constants ($h_1$, $h_2$, etc.)[30, 31] can be separated parametrically by isotopic analysis.[20, 24] In the present model, $H_{JT}$ is absorbed perturbatively by the rotational term of the effective Hamiltonian (Eq. 15) if different rotational constants are used for the $\Gamma = \pm 1$ or $\wp = \pm 1$ states, respectively. Hence this term is not included in Eq. 15.

### 5.2. Separation and mixing of electronic states

Using the Hund's case (a) coupling scheme, the nonrotating-molecule Hamiltonian is diagonal with respect to $J$ and $S$, and has the form:

$$H_q + H_{SO} = \begin{array}{cc} |\Omega|=3/2 & |\Omega|=1/2 \end{array} \frac{1}{2}\begin{pmatrix} a\zeta_e d & \Delta E_0 \\ \Delta E_0 & -a\zeta_e d \end{pmatrix}. \tag{33}$$

where $\Omega = \Lambda + \Sigma$. The energy separation between the rotationless levels of the two coupled states is therefore:

$$\Delta E = \sqrt{(a\zeta_e d)^2 + \Delta E_0^2} = \sqrt{(a\zeta_e d)^2 + (\varepsilon_2 + \Delta ZPE)^2}. \tag{34}$$

For certain molecules, one or more of the three parameters on the right side of the equation ($a\zeta_e d$, $\varepsilon_2$, $\Delta ZPE$) may vanish. For instance, for the normal and the fully deuterated isotopologues of the methoxy radical (CH$_3$O and CD$_3$O),[44, 57] only the effective SO constant $a\zeta_e d$ is nonzero, while for its asymmetrically deuterated isotopologues (CH$_2$DO and CHD$_2$O), $\Delta ZPE$ is also nonzero.[23, 24] For asymmetrically deuterated cyclopentadienyls,[20] both $a\zeta_e d$ and $\varepsilon_2$ vanish, and the separation between the two lowest electronic states is solely due to $\Delta ZPE$. For pJT-active free radicals, usually all three parameters are nonzero. Additionally, the Coriolis interaction further separates the two electronic states of a *rotating* molecule, although to a much less extent. The shift of individual rotational levels due to $H_C$ depends on their rotational quantum numbers $N$.

Readers are referred to Refs. [12, 17, 24, 25, 27, 58, 59] for detailed discussion on energy level structure for different coupling strengths with varying $a\zeta_e d$ and $\Delta E_0$. Although discussion therein is limited to $C_s$ or $C_{2v}$ molecules, the general conclusions hold. (Note that in Refs. [12, 17], $a\zeta_e d$ and $\varepsilon_2$ are denoted as $A_{SO}$ and $\rho$, respectively.) Quantitative analysis of energy level structure of free radicals with lower symmetry using real-world examples would be beneficial but is best dealt with separately in future publications.

### 5.3. SR constants

The SR term in the effective Hamiltonian deserves further elaboration. In general SR interaction can be treated as a perturbation to the rotational Hamiltonian.[47] The first-order perturbation contribution ($\varepsilon_{\alpha\beta}^{(1)}$) arises from the interaction of the magnetic field generated by the rotation of the nuclei with the electron spin magnetic moment. The second-order perturbation contribution ($\varepsilon_{\alpha\beta}^{(2)}$) arises from the cross term of the SO and the Coriolis interactions. Liu and Miller further demonstrated that, for vibronically coupled states, cross terms of various Hamiltonians make contribution to effective molecular constants such as SR constants, as well as create new constants in certain cases.[60, 61] Such contributions have been used to explain the unusually large SR constant $\varepsilon_{zz}$ of the methoxy radical.[62]

It is well known that the major contribution to SR constants is the seconder-order perturbation. When the energy separation between the interacting electronic states is large compared to the SO and Coriolis interaction, i.e., $\Delta E_0 \gg a\zeta_e d$ and $\Delta E_0 \gg \mu_{z'z'}\zeta_t$, the second-order contribution to the $n^{\text{th}}$ energy level can be calculated using perturbation theory:[63-67]

$$\varepsilon_{\alpha\beta} \approx \varepsilon_{\alpha\beta}^{(2)} = -\sum_{\delta}\sum_{n'\neq n} \frac{\langle n'|\mu_{\alpha\delta}(L_\delta + G_\delta)|n\rangle\langle n|aL_\beta|n'\rangle + \langle n'|aL_\beta|n\rangle\langle n|\mu_{\alpha\delta}(L_\delta + G_\delta)|n'\rangle}{E_n - E_{n'}}, \tag{35}$$

where $\alpha$, $\beta$, $\delta$ are Cartesian coordinates. $E_n - E_{n'}$ is the energy separation between the $n^{\text{th}}$ and the $n'^{\text{th}}$ levels. In IAS, contribution of $L_{x'}$ and $L_{y'}$ can be neglected. Eq. (35) reduces to:

$$\begin{aligned}\varepsilon_{z'z'} &= -2\sum_{\delta=x',y',z'}\sum_{n'\neq n}\frac{\langle n'|aL_{z'}|n\rangle\langle n|\mu_{z'\delta}(L_{z'} + G_{z'})|n'\rangle}{E_n - E_{n'}} \\ &= -2\sum_{\delta=x',y',z'}\mu_{z'\delta}\sum_{n'\neq n}\frac{(a\zeta_e d)_{n'n}(\zeta_t)_{nn'}}{E_n - E_{n'}}\end{aligned}, \tag{36}$$

where $(a\zeta_e d)_{n'n} = \langle n'|aL_{z'}|n\rangle$ and $(\zeta_t)_{nn'} = \langle n|(L_{z'} + G_{z'})|n'\rangle$. In Eq. (35) and (36), it is assumed that the two coupled electronic states have identical principal axes and moments of inertia. In the case of two nearly degenerate electronic states, Eq. (36) further reduces to:

$$\varepsilon_{z'z'} = \varepsilon'\sum_{\delta=x',y',z'}\mu_{z'\delta}, \tag{37}$$

where $\varepsilon'$ is the mass-independent SR constant:[67, 68]

$$\varepsilon' = \pm \frac{2a\zeta_e d\zeta_t}{\Delta E}$$

(38)

The " + " and " – " signs in Eq. (38) are for the lower and the upper states, respectively.

SR constants are usually given in the PAS. The SR tensor in the PAS can be calculated using $\varepsilon'$ and through a unitary transformation from the IAS to the PAS: $\boldsymbol{\varepsilon} = \boldsymbol{\mu} U \begin{pmatrix} \varepsilon' & 0 & 0 \\ 0 & 0 & 0 \\ 0 & 0 & 0 \end{pmatrix} U^{-1}$. Since this is in the PAS, $\boldsymbol{\mu}$ is diagonal and $\mu_{\alpha\beta} = 2B_\alpha \delta_{\alpha,\beta}$. The result is:

$$\begin{aligned}
\varepsilon_{zz} &= 2B_z \varepsilon' \cos^2\theta \\
\varepsilon_{xx} &= 2B_x \varepsilon' \sin^2\theta \cos^2\phi \\
\varepsilon_{yy} &= 2B_y \varepsilon' \sin^2\theta \sin^2\phi \\
(\varepsilon_{zx} + \varepsilon_{xz})/2 &= (B_z + B_x)\varepsilon' \sin\theta \cos\theta \cos\phi \\
(\varepsilon_{xy} + \varepsilon_{yx})/2 &= (B_x + B_y)\varepsilon' \sin^2\theta \sin\phi \cos\phi \\
(\varepsilon_{yz} + \varepsilon_{zy})/2 &= (B_y + B_z)\varepsilon' \sin\theta \cos\theta \sin\phi
\end{aligned}$$ (39)

Modern quantum chemical calculations can predict to high precision the geometry and, hence, rotational constants of molecules.[69] Although calculating the effective SO constant[69] and the Coriolis constant *ab initio* is still a challenging task, they often can be determined in vibronic analysis.[5] SR constants can therefore be calculated using Eqs. (38) and (39), and used as initial values in simulating rotationally resolved spectra. In addition, these equations can be employed to derive the relation of SR constants upon isotopic or chemical substitution, [67, 68, 70] and to predict

SR constants of a substitution using a reference molecule with known SR constants.[27, 71, 72] In such calculations, the mass-independent SR constant $\varepsilon'$ is assumed to be invariant upon substitution.

When $\Delta E_0$ is comparable or smaller than $a\zeta_e d$ or $\mu_{z'z'}\zeta_t$, the values of SR constants cannot be calculated perturbatively.[25] SO and Coriolis constants need to be used in spectral simulation with the effective Hamiltonian.

### 5.4. Vibronic interactions with other states and extension of the Hamiltonian matrix

In the coupled-states Hamiltonian, the SO and Coriolis interactions between the two nearly degenerate energy levels are included explicitly. In the absence of other interacting states, all SR constants in the effective Hamiltonian are assumed to vanish. However, for open-shell molecules in nearly degenerate states, this is seldom the case because the vibrational ground level of one electronic state can interact with vibrational levels of the other electronic state.[60, 61] Interactions with other levels can be taken into account by either of the following approaches:

(i) One may increase the size of the Hamiltonian and include the SO and Coriolis interactions between all vibronic levels explicitly. The new Hamiltonian terms have the same form as Eqs. (23) and (27), although the values of the molecular constants ($a\zeta_e d$ and $\zeta_t$) are in general different.

(ii) Alternatively, the interaction with a third (or more) level can be treated perturbatively by including the SR Hamiltonian term. In this case, SO and Coriolis terms are to describe coupling between the two nearly degenerate states, whereas the SR term describes coupling to the third state phenomenologically. In general, the two nearly degenerate states are expected to have different SR constants.

## 5.5. Correlation of molecular constants

In the absence of interaction with a third state, either the SR tensor or the combination of SO and Coriolis constants can reproduce the SR splitting. As a result, strong correlation between these molecular constants may occur in the following situations:

1) Only one of the two nearly degenerate states is accessed in high-resolution spectroscopic measurements, i.e. with its rotational and fine structure resolved;

2) A third (or more) state is coupled to either or both of the two states *via* SO and Coriolis interactions, and this third state is not accessed in high-resolution spectroscopic measurements.

Especially, in the case of $C_s$ molecules, it has been found that the SO constant is strongly coupled to $\varepsilon_{z'z'}$, whereas the Coriolis constant is coupled to $\varepsilon_{x'x'}$ and $\varepsilon_{y'y'}$.[27] Moreover, in the forgoing Case (1), the SO and Coriolis constants are also coupled to each other (see Section 7). Breaking the correlation requires (i) spectroscopic access to all interacting states with rotational and fine structure resolution, or (ii) accurate determination of one or more of the correlated constants by computation (see Section 7).

## 6. Transition intensities and selection rules

Transition intensity from an initial energy level $i$ to a final energy level $f$ can in general be written as:

$$I(i;f) = C(N_i - N_f)S(i;f), \tag{40}$$

where $C$ is an instrumental constant, $N_i$ and $N_f$ are the populations of the $i$ and $f$ states, respectively, and $S(i;f)$ is the line strength of the transition. For an electric dipole transition, $S(i;f)$ can be written as:

$$S(i;f) = 3\sum |\mu(i;f)|^2 = 3\sum |\langle i|\mu_Z|f\rangle|^2, \tag{41}$$

where $\mu_Z$ is the space-fixed $Z$-component of the electric dipole operator of the molecule, and $\mu(i;f) = \langle i|\mu_Z|f\rangle$ is the transition dipole moment between the two states. Summation in Eq. (41) is over the space-fixed projection quantum numbers in both the $i$ and $f$ states. In the present work, the projection of $\bm{J}$ onto the $Z$-axis is labelled $M$ with $M = -J, -J+1\ldots J-1, J$. In Eq. (41) an isotropic or "natural" excitation is assumed,10 hence the factor of 3 before the summation notation. The operator $\mu_Z$ corresponds to the $0^{\text{th}}$ component of a first-rank irreducible tensor operator in a space-fixed coordinate system, which is related to the corresponding components, $\mu_q$ with $q = 0, \pm 1$, in the molecule-fixed system by:

$$\mu_Z = \mu_0^1 = \sum_{q=0,\pm 1} D_{0q}^{1*} T_q^1(\mu), \tag{42}$$

where $D_{0q}^1$ is the rotational matrix relating the molecule- and space-fixed systems, and $T_q^1(\mu)$ is the first-rank irreducible tensor of the electric dipole operator. In a spherical basis set, the components of $T_q^1(\mu)$ are:[73]

$$\begin{aligned} T_0^1(\mu) &= \mu_{z'} \\ T_{\pm 1}^1(\mu) &= \mp \tfrac{1}{\sqrt{2}}(\mu_{x'} \pm i\mu_{y'}) \end{aligned}. \tag{43}$$

In the equations above, the IAS is adopted. Substitution of Eq. (42) into Eq. (41) gives:

$$S(i;f) = 3 \sum_{M,M'} \left| \sum_{q=0,\pm 1} \langle i | D_{0q}^{1*} T_q^1(\mu) | f \rangle \right|^2 . \tag{44}$$

As any eigenfunction ($|i\rangle$ or $|f\rangle$) is a linear combination of basis functions, it is sufficient to calculate the matrix elements of $\mu_q^1$ in the initial- and final-state basis functions and perform the appropriate summation over the molecule-fixed quantum numbers. One further utilizes the fact that Hamiltonian matrices are block-diagonalized with respect to good quantum numbers ($J$ and $S$) so that only summation over the bad quantum numbers ($P$ and $\Sigma$ in the case (a) basis sets; $N$ and $K$ in the case (b) basis sets, $\Gamma$ and $\wp$ in both cases) is necessary. The overall formula for calculating the line strength is therefore:

$$S(i;f) = 3 \sum_{M,M'} \left| \sum_{q=0,\pm 1} \sum_{i,f} a_i a_f \langle \Phi_i | D_{0q}^{1*} T_q^1(\mu) | \Phi_f \rangle \right|^2 , \tag{45}$$

where $|\Phi_i\rangle$ and $|\Phi_f\rangle$ are basis functions of the initial and the final states, respectively. The last summation in Eq. (45) is over all basis functions with the same good quantum numbers, and $a_i$ and $a_f$ are the expansion coefficients of the eigenfunctions, which can be extracted in the diagonalization of the Hamiltonian matrices.

In the present work, we consider only transitions between an isolated, orbitally nondegenerate electronic state with $\Lambda = 0$, and a pair of nearly degenerate states, modelled by the coupled-state Hamiltonian. For the $\Lambda = 0$ state, an orbital symmetrization like those in Eqs. (5) and (6) loses its meaning. Transition intensity formulae in the orbitally symmetrized basis sets are therefore identical to the well-studied transitions between isolated electronic states *except* that transitions involving the $\Lambda = +1$ and the $\Lambda = -1$ orbitals have different *but related* electronic transition dipole moments. Combining these two types of transitions with the symmetry relations, line strength formulae $S(i;f)$ and selection rules in the orbitally symmetrized basis sets can be derived as shown in the Supplementary Materials (see Sections S.8 and S.9).

We now discuss the transition intensities in the fully symmetrized basis sets. Similar to treatment of the nearly degenerate electronic state, symmetrized spin-ro-orbital basis functions can be constructed for the $\Lambda = 0$ state as:

(i) Fully symmetrized case (a) basis set:

$$\left|J,\bar{P},S,\bar{\Sigma},\wp\right\rangle = \tfrac{1}{\sqrt{2}}\left[\left|\Lambda=0,J,P,S,\Sigma\right\rangle + \wp(-1)^{J-P+S-\Sigma}\left|\Lambda=0,J,-P,S,-\Sigma\right\rangle\right]; \qquad (46)$$

(ii) Fully symmetrized case (b) basis set:

$$\left|J,N,\bar{K},S,\wp\right\rangle = \tfrac{1}{\sqrt{2}}\left[\left|\Lambda=0,J,N,K,S\right\rangle + \wp(-1)^{N-K}\left|\Lambda=0,J,N,-K,S\right\rangle\right]. \qquad (47)$$

Hamiltonian matrices of the $\Lambda = 0$ state can be computed using the same methods in Sections 3 and 4. Due to the absence of $H_q$, $H_{SO}$, $H_C$, as well as the JT terms, in the effective Hamiltonian for the $\Lambda = 0$ state, an *artificial* degeneracy of the $\wp = \pm 1$ levels is introduced. A degeneracy of the $\Sigma = \pm 1/2$ levels (in case (a) basis sets) or $J = N \pm 1/2$ levels (in case (b) basis sets) also exists if all SR constants vanish.

Formulae of line strength in the fully symmetrized basis sets can be derived based on that for an asymmetric top and with symmetry relation between the $|\Lambda = \pm 1, J, \pm P, S, \pm \Sigma\rangle$ or $|\Lambda = \pm 1, J, N, \pm K, \Sigma, S\rangle$ basis functions employed. Details of the calculations are given in the Supplementary Materials (see Sections S.10 and S.11). Here in the main text, we summarize the final formulae for the line strength in all four basis sets: (Primed are $\Lambda = 0$ state quantum numbers.)

(i) In the orbitally symmetrized case (a) basis set:

$$S(\Gamma',J',P',S',\Sigma';\Gamma,J,P,S,\Sigma)$$

$$= \frac{1}{2}\delta_{S,S'}(2J+1)(2J'+1)\left|\sum_{q=0,\pm 1}\left\{\sum_{\Gamma}\sum_{P,P'}\sum_{\Sigma,\Sigma'}a_i a_f \delta_{\Sigma,\Sigma'}(-1)^{J'+P-1}\left[\begin{pmatrix} J & 1 & J' \\ P & q & -P' \end{pmatrix} + (-1)^q \Gamma \begin{pmatrix} J & 1 & J' \\ P & -q & -P' \end{pmatrix}\right]\right\}\mathbf{M}_q\right|^2. \quad (48)$$

(ii) In the orbitally symmetrized case (b) basis set:

$$S(\Gamma',J',N',K',S';\Gamma,J,N,K,S)$$

$$= \frac{1}{2}\delta_{S,S'}(2J+1)(2J'+1)\left|\sum_{\Gamma}\sum_{N,N'}(-1)^{J'+N+S+1}\begin{pmatrix} N & J & S \\ J' & N' & 1 \end{pmatrix}\tilde{S}^{1/2}(J',N',K',S';J,N,K,S)\right|^2, \quad (49)$$

where

$$\tilde{S}^{1/2}(J',N',K',S';J,N,K,S)$$

$$= (2N+1)^{1/2}(2N'+1)^{1/2}\sum_{q=0,\pm 1}\left\{\sum_{K,K'}a_i a_f (-1)^{N-K'-1}\left[\begin{pmatrix} N & 1 & N' \\ K & q & -K' \end{pmatrix} + (-1)^q \Gamma \begin{pmatrix} N & 1 & N' \\ K & -q & -K' \end{pmatrix}\right]\right\}\mathbf{M}_q. \quad (50)$$

(iii) In the fully symmetrized case (a) basis set:

$$S(J',\bar{P}',S',\bar{\Sigma}',\wp';J,\bar{P},S,\bar{\Sigma},\wp)$$
$$= \frac{1}{2}\delta_{S,S'}(2J+1)(2J'+1)\left|\sum_{q=0,\pm 1}\left[\sum_{\wp,\wp'}\sum_{\bar{P},\bar{P}'}\sum_{\bar{\Sigma},\bar{\Sigma}'}a_i a_f \tilde{F}^{1/2}(J',\bar{P}',S',\bar{\Sigma}',\wp';J,\bar{P},S,\bar{\Sigma},\wp)\right]\mathbf{M}_q\right|^2, \quad (51)$$

where

$$\tilde{F}^{1/2}(J',\bar{P}',S',\bar{\Sigma}',\wp';J,\bar{P},S,\bar{\Sigma},\wp) = \begin{cases} (-1)^{J'+\bar{P}-1}\delta_{\wp,-\wp'}\begin{pmatrix} J & 1 & J' \\ -\bar{P} & q & \bar{P}' \end{pmatrix} & \text{if } \bar{\Sigma}' = \bar{\Sigma} \\ s'\wp'(-1)^{J'+\bar{P}-1}\delta_{\wp,-\wp'}\begin{pmatrix} J & 1 & J' \\ \bar{P} & q & \bar{P}' \end{pmatrix} & \text{if } \bar{\Sigma}' = -\bar{\Sigma} \end{cases}. \quad (52)$$

(iv) In the fully symmetrized case (b) basis set:

$$S(J',N',\bar{K}',S',\wp';J,N,\bar{K},S,\wp) =$$
$$= \frac{1}{4}\delta_{S,S'}(2J+1)(2J'+1)\left|\sum_{\wp \wp'}\sum_{N,N'}(-1)^{J'+N+S+1}\begin{pmatrix} N & J & S \\ J' & N' & 1 \end{pmatrix}\tilde{S}^{1/2}(J',N',K',S';J,N,K,S)\right|^2, \quad (53)$$

where

$$\tilde{S}^{1/2}(J',N',\bar{K}',S';J,N,\bar{K},S)$$
$$= 2\delta_{\wp,-\wp'}[1+s'\wp'](2N+1)^{1/2}(2N'+1)^{1/2}\sum_{q=0,\pm 1}\left\{\sum_{K,K'}a_i a_f (-1)^{N-K'-1}\begin{pmatrix} N & 1 & N' \\ \bar{K} & q & -\bar{K}' \end{pmatrix}\right\}\mathbf{M}_q. \quad (54)$$

In the equations above, $\mathbf{M}_q = \langle \Lambda'=0|T_q^1(\mu)|\Lambda=+1\rangle = (-1)^q \langle \Lambda'=0|T_{-q}^1(\mu)|\Lambda=-1\rangle$ are the components of the electronic transition dipole moment between the two states in the spherical basis set.

All four line strength formulae (Eq. 48, 49, 51, (53)) are equivalent to each other. They reveal different aspects of the transition types. Selection rules in the four basis sets can be determined from the line strength formulae by inspecting the delta functions and using the triangular conditions of the 3-*j* symbols. Collectively, the selection rules are:

$J' = J, J \pm 1$;

$P' = P, P \pm 1$ for $q = 0, \pm 1$;

If $\bar{\Sigma}' = \bar{\Sigma}$, $\bar{P}' = \bar{P}, \bar{P} \pm 1$ for $q = 0, \pm 1$; If $\bar{\Sigma}' = -\bar{\Sigma}$, $\bar{P}' = -\bar{P}, -\bar{P} \pm 1$ for $q = 0, \pm 1$;

$S' = S$;

$\Sigma' = \Sigma$;

$N' = N, N \pm 1$;

$K' = K, K \pm 1$ for $q = 0, \pm 1$;

$\bar{K}' = \bar{K}, \bar{K} \pm 1$ for $q = 0, \pm 1$;

$\wp' = -\wp$.

(See Sections S.4-S.7 for details.)

Due to vibronic interactions and, hence, breakdown of the Born-Oppenheimer approximation, the allowed transition types, i.e., the orientation of the transition dipole moment, is determined by the vibronic symmetry, instead of the orbital symmetry, of the initial and the final states. For transitions between vibrational ground levels of an isolated $\Lambda = 0$ state and a pair of coupled $\Lambda = \pm 1$ states, only perpendicular transitions ($\mathbf{M}_q$ with $q = \pm 1$) are allowed. Inspection of the term in the square brackets in Eqs. (48) or (50) with $q = \pm 1$ applied suggests that for transitions from or to the $\Gamma = +1$ basis functions, i.e., the $A'$ state, the electronic transition dipole is along the $x'$ axis, whereas it is along the $y'$ axis for transitions from or to the $\Gamma = -1$ basis functions, i.e., the $A''$ state. (See Section S.4.) If the fully symmetrized basis sets are used, allowed transition types can be determined from the spin-ro-orbital symmetry ($\wp$) and the spin-rotational symmetry ($s$) of the basis functions using the relation $\Gamma = s\wp$. The transition dipole moment can be readily converted from the IAS to PAS through a unitary transformation of coordinate systems.

Compared to isolated electronic states, rotational energy levels of $A'$ and $A''$ basis functions are mixed in coupled electronic state by the SO interaction and, to a much less extent, the Coriolis interaction. Such mixing leads to transitions that are not predicted by the isolated-states model.[27]

## 7. Examples: High-resolution spectra of the trans (T) and gauche (G) conformer of the 1-propoxy radical

Methoxy, the smallest alkoxy radical, is a prototypical molecule whose ground electronic ($\tilde{X}^2E$) state is subject to both JT and SO interactions. Other alkoxy radicals can be regarded as alkyl substitutions of methoxy. Except for those that retain the $C_{3v}$ symmetry, e.g., $t$-butoxy, alkyl substitution leads to nonzero $\Delta E_0$ and $\Delta ZPE$ between the two vibrational ground levels, and the $\tilde{X}^2E$ state of methoxy is split into two electronic states, which are of $A'$ and $A''$ symmetry if a $C_s$ plane is maintained. Previously, the rotational and fine structure of the two lowest electronic states of isopropoxy ($\tilde{X}^2A'$ and $\tilde{A}^2A''$) and cyclohexoxy ($\tilde{X}^2A''$ and $\tilde{A}^2A'$) was investigated using a reduced version of the present effective Hamiltonian that is specifically for $C_s$ molecules. In this section, we employ the effective Hamiltonian developed in the present work to simulate $\tilde{B} \leftarrow \tilde{X}$ transitions of both trans [T ($C_s$)] and gauche [G ($C_1$)] conformers of the 1-propoxy radical.[71] The $\tilde{B}$ state is a well separated state with $\Lambda = 0$ character.

For all alkoxy radicals, the unpaired electron is localized at the oxygen atom. Therefore one expects that $L$ and $G$ are oriented along the CO bond, which is taken as the $z'$ axis. It is also assumed that the $x'$ axis is within the $OC_\alpha C_\beta$ plane, while the $y'$ axis is perpendicular to it. (See Figure 3). As mentioned before (Section 2) and will be demonstrated later, these assumptions are only for the sake of convenience of discussion. They do not affect the final values of molecular constants extracted from spectroscopic fitting. From a practical point of view, these assumptions, although not always precise, help to determine the initial values for $\theta$ and $\phi$ used in the fitting because the geometry of molecules and, hence, the orientation of the CO bond, can be calculated *ab initio* with high precision.

Spectra of the $\tilde{B} \leftarrow \tilde{X}$ transitions simulated using the coupled-states Hamiltonian are compared with experimental ones in Figure 4. All simulations as well fitting are done using the SpecView software[74] which the new spectroscopic models are incorporated in. The fit values for molecular constants are listed in Table 1. Compared to Fit 1 (with the isolated-states model), Fit 2 (with the coupled-states model) uses one fewer *fit* molecular constants for the T conformer, and two fewer *fit* molecular constants for the G conformer. This is because in the coupled-states model, the SR splitting is no longer reproduced using four (for the T conformer) or fix (for the G conformer) SR constants, but rather the SO constant ($a\zeta_e d$), the Coriolis constant ($\zeta_t$), and the orientation angle(s) of $L$ and $G$ ($\theta$ for the T conformer; $\theta$ and $\phi$ for the G conformer). Another more significant advantage of Fit 2 is that SO constants, Coriolis constant, and orientation angles can be calculated or estimated with higher precision than SR constants. Note that in the present work the $I^r$ representation for an asymmetric top is used so that $z=a$, $x=b$, $y=c$, and $\theta$ and $\phi$ are determined using aforementioned definitions (see Figure 1a).

Due to the fact that only one of the two nearly degenerated electronic states, the lower-energy $\tilde{X}$ state, is accessed experimentally in each of the two cases (T and G conformers), the SO and Coriolis constants are strongly coupled to each other in the fit (see Section 5.5) and cannot be determined independently with high precision. Hence the large error bars for these two constants in Fit 2. Furthermore, the fit values of the SO and Coriolis constants may be contaminated by correlation with the SR constants, which are fixed to zero in Fit 2. (See Section 5.3) Fortunately, for both conformers, the energy separation between the vibrational ground levels of the $\tilde{X}$ and the $\tilde{A}$ state ($\Delta E^{\tilde{A}-\tilde{X}}$) is large so that the vibrational ground level of the $\tilde{X}$ state is well separated from vibrational levels of the $\tilde{A}$ state. Coupling between these states and, hence, the contamination to $a\zeta_e d$ and $\zeta_t$ of the $\tilde{X}$ state, are expected to be small. Nevertheless, the error bars quoted in Table 1 represent the standard deviations of the parameters determined in fitting, not the real uncertainties of these parameters, nor the estimated deviation from their real values. A precise determination of these two constants requires inclusion of experimental transitions involving the $\tilde{A}$ state into the fitting.

In the absence of experimental data involving the $\tilde{A}$ state, it is necessary to estimate the value of either of these two constants ($a\zeta_e d$ and $\zeta_t$). For 1-propoxy, as well as other primary alkoxy radicals, the pJT coupling between the $\tilde{X}$ and $\tilde{A}$ electronic states is weak compared to their energy separation ($\Delta E^{\tilde{A}-\tilde{X}}$). Under this condition, the electronic quenching factor of the SO constant ($\zeta_e$) and the Coriolis constant ($\zeta_t$) are related to each other by the following equation:[5]

$$\zeta_t \approx \zeta_e + \sum_i l_i \zeta_i , \qquad (55)$$

where $i$ denotes different vibrational modes. For the vibrational ground level, $l_i = 0$. Thus $\zeta_t \approx \zeta_e$, i.e., it is assumed that $(L_{z'} + G_{z'})$ and $L_{z'}$ have the same expectation value in the $\tilde{X}$-state vibronic eigenfunction. One may take the SO splitting of the OH radical[43] (-139 cm$^{-1}$) as the estimated value for $a$.[5, 42] Coupled-cluster calculations predict that $a\zeta_e$ is 131 cm$^{-1}$ for both conformers of 1-propoxy.[75] $\zeta_e$ is therefore estimated to be 0.942, and so is $\zeta_t$ based on the discussion above. Another fit (Fit 3) using the coupled-states model with $\zeta_t$ fixed to 0.942 was carried out for each conformer. The values for molecular constants determined in Fit 3 are also listed in Table 1.

All three fits give close values for rotational constants, which are predicted by density functional theory (DFT) calculations with high precision. Molecular constants determined in Fit 2 and Fit 3 have close values except for $a\zeta_e d$ and $\zeta_t$. DFT calculations do not predict the SR constants with good precision. For both conformers, the fit values of $\theta$ using the coupled-states model are very close to those predicted by DFT calculated ones. However, the fit values of $\phi$ for the G-conformer deviate from the calculated value by ~36%. Inspection of the molecular geometry (see Figure 3b) suggests that this deviation is due to the influence of the terminal carbon atom, which causes $\boldsymbol{L}$ and $\boldsymbol{G}$ to tilt out of the OC$_\alpha$C$_\beta$ plane, viz., the $z'x'$ plan. A better prediction of $\phi$ as well as $\theta$ could be reached using the calculated orientation of the molecular orbitals instead of the CO bond orientation. Allowing $\boldsymbol{L}$ and $\boldsymbol{G}$ to have different orientation angles doesn't increase the quality of the fitting, suggesting that it is reasonable to assume that these two angular momenta have the same orientation.

## 8. Conclusions and outlook

Spin-ro-vibronic structure of molecules in orbitally degenerate electronic states including RT and JT-active molecules has been extensively studied. Less is known about rotational structure of nonlinear polyatomic molecules in nearly degenerate states. In the case of free radicals, the unpaired electron further complicates energy levels by introducing SO and SR splitting. The study on free radicals in nearly degenerate states provides a promising avenue of research which may bridge the gap between symmetry-induced degenerate states and the Born-Oppenheimer limit of unperturbed electronic states.

In the present work, we have developed a new spectroscopic model that can be used to analyze and understand the spin-ro-vibronic structure of molecules in nearly degenerate states, and to simulate their experimentally obtained high-resolution spectra. The spectroscopic model is not limited by symmetry requirements although certain assumptions have been made for the sake of convenience of discussion. The two major mechanisms that make contributions to the separation of the electronic states: (i) the SO interaction, and (ii) the combined nonrelativistic effect of difference potential and ZPE difference between the two electronic states, can now be determined independently.

The coupled-states Hamiltonian can be expanded to deal with multiple interacting states. (i) It is straightforward to include more electronic or vibronic states for molecules with double orbital (near) degeneracy. One simply increases the number of Hamiltonian blocks (see Figure 2) and computes the matrix elements using the formulae in the Appendices. (ii) If the degree of orbital degeneracy is three or higher, new symmetrized basis sets that transform according to the irreducible representations of the point group need to be constructed. The Hamiltonian matrix can be constructed according to the topological relations between the eigenstates of the nonrotating molecule.[76-78] Action rules of the rotational and spin angular momentum operators remain the same as those in Sections S.2 and S.3.

The present paper uses the $\tilde{B} \leftarrow \tilde{X}$ transition of the two conformers of the 1-propoxy radical as an example. The $\tilde{X}$ state is coupled to the $\tilde{A}$ state through SO and pJT interactions. Unfortunately, limited by Boltzmann distribution at low temperature under jet-cooled conditions, the $\tilde{A}$ state of alkoxy radicals has not been investigated with rotational resolution except for cyclohexoxy.[27] Accurate determination of the SO and Coriolis constants is therefore hindered by correlation of these two parameters although one of them ($\zeta_t$) can be estimated on the basis of electronic structure calculations.

Correlation of molecular constants can be removed when both of the two coupled electronic states are accessed with rotational and fine-structure resolution. In laser spectroscopy, this is easy to realize when both states are electronic excited states, such as in the case of alkyl-substituted organometallic monomethyl and monomethoxide radicals,[79-86] e.g., $CaCH_2CH_3$, $CaCH(CH_3)_2$, $CaOCH_2CH_3$, and $CaOCH(CH_3)_2$. Experimentally obtained high-resolution laser spectra of these molecules and their analysis with the present spectroscopic model would provide valuable information about SO and vibronic interactions, as well as related effects in free radicals. SO and Coriolis constants determined in simulating the rotational and fine structure can aid in vibronic analysis[5] and interpretation of effective molecular constants such as the SR constants. Furthermore, the two families of molecules mentioned above have been proposed recently as candidates for future laser cooling of polyatomic molecules.[87-89] High-resolution spectroscopic investigation of these molecules therefore has important significance to atomic, molecular and optical physics.

**Acknowledgements:**

The author is grateful to Dr. Terry A. Miller (Ohio State) for his continuous encouragement and fruitful discussion, and to Dr. Lan Cheng (Johns Hopkins) for sharing preliminary results from his coupled cluster calculations of $\zeta_e$'s of alkoxy radicals. This work was supported by the National Science Foundation under grant number CHE-1454825.

**References:**

1. R. Renner, Z. Phys. **92**, 172 (1934).
2. J. M. Brown, in *Computational molecular spectroscopy*, edited by P. Jensen and P. R. Bunker (John Wiley & Sons, 2000).
3. H. A. Jahn and E. Teller, Proc. R. Soc. London, Ser. A **161**, 220 (1937).
4. I. B. Bersuker, *The Jahn-Teller Effect*. (Cambridge University Press, 2006)
5. T. A. Barckholtz and T. A. Miller, Int. Rev. Phys. Chem. **17**, 435 (1998).
6. G. Herzberg, *Molecular Spectra And Molecular Structure I: Spectra of Diatomic Molecules.* (D. van Nostrand Company, 1945)
7. G. Herzberg, *Molecular Spectra And Molecular Structure II: Infrared and Raman Spectra of Polyatomic Molecules*. (D. van Nostrand Company, 1945)
8. E. Hirota, *High-Resolution Spectroscopy of Transient Molecules*. (Springer Verlag, 1985)
9. P. D. A. Mills, C. M. Western and B. J. Howard, J. Phys. Chem. **90**, 3331 (1986).
10. W. M. Fawzy and J. T. Hougen, J. Mol. Spectrosc. **137**, 154 (1989).
11. W. H. Green and M. I. Lester, J. Chem. Phys. **96**, 2573 (1992).
12. M. D. Marshall and M. I. Lester, J. Chem. Phys. **121**, 3019 (2004).
13. C. S. Brauer, G. Sedo, E. M. Grumstrup, K. R. Leopold, M. D. Marshall and H. O. Leung, Chem. Phys. Lett. **401**, 420 (2005).
14. M. D. Marshall and M. I. Lester, J. Phys. Chem. B **109**, 8400 (2005).
15. Y. Ohshima, K. Sato, Y. Sumiyoshi and Y. Endo, J. Am. Chem. Soc. **127**, 1108 (2005).
16. S. Wu, G. Sedo and K. R. Leopold, J. Mol. Spectrosc. **253**, 35 (2009).
17. G. E. Douberly, P. L. Raston, T. Liang and M. D. Marshall, J. Chem. Phys. **142**, 134306 (2015).
18. F. J. Hernandez, J. T. Brice, C. M. Leavitt, G. A. Pino and G. E. Douberly, J. Phys. Chem. A **119**, 8125 (2015).
19. A. Carrington, H. C. Longuet-Higgins, R. E. Moss and P. F. Todd, Mol. Phys. **9**, 187 (1965).
20. L. Yu, D. W. Cullin, J. M. Williamson and T. A. Miller, J. Chem. Phys. **98**, 2682 (1993).
21. H. J. Wörner and F. Merkt, J. Chem. Phys. **126**, 154304 (2007).
22. M. Grütter, H. J. Wörner and F. Merkt, J. Chem. Phys. **131**, 024309 (2009).
23. D. Melnik, J. Liu, R. F. Curl and T. A. Miller, Mol. Phys. **105**, 529 (2007).
24. D. G. Melnik, J. Liu, M.-W. Chen, T. A. Miller and R. F. Curl, J. Chem. Phys. **135**, 094310 (2011).
25. J. Liu, D. Melnik and T. A. Miller, J. Chem. Phys. **139**, 094308 (2013).
26. R. Chhantyal-Pun, M. Roudjane, D. G. Melnik, T. A. Miller and J. Liu, J. Phys. Chem. A **118**, 11852 (2014).
27. J. Liu and T. A. Miller, J. Phys. Chem. A **118**, 11871 (2014).
28. L. Zu, J. Liu, G. Tarczay, P. Dupré and T. A. Miller, J. Chem. Phys. **120**, 10579 (2004).
29. A. Carrington, A. R. Fabris, B. J. Howard and N. J. D. Lucas, Mol. Phys. **20**, 961 (1971).
30. J. T. Hougen, *The Calculation of Rotational Energy Levels and Rotational Line Intensities in Diatomic Molecules*. (National Bureau of Standards, Washington D.C., 1970)
31. J. T. Hougen, J. Mol. Spectrosc. **81**, 73 (1980).
32. Y. Endo, S. Saito and E. Hirota, J. Chem. Phys. **81**, 122 (1984).
33. J. T. Hougen, J. Chem. Phys. **37**, 1433 (1962).
34. J. T. Hougen, J. Chem. Phys. **39**, 358 (1963).
35. H. C. Longuet-Higgins, Mol. Phys. **6**, 445 (1963).
36. K. Kawaguchi, E. Hirota, T. Ishiwata and I. Tanaka, J. Chem. Phys. **93**, 951 (1990).
37. I. J. Kalinovski, Ph.D. Thesis, University of California, Berkeley, 2001.
38. J. M. Brown, Mol. Phys. **20**, 817 (1971).

39. J. K. G. Watson, J. Mol. Spectrosc. **50**, 281 (1974).
40. H.-B. Qian, D. Seccombe and B. J. Howard, J. Chem. Phys. **107**, 7658 (1997).
41. D. S. McClure, J. Chem. Phys. **20**, 682 (1952).
42. M. Grütter, X. Qian and F. Merkt, J. Chem. Phys. **137**, 084313 (2012).
43. J. P. Maillard, J. Chauville and A. W. Mantz, J. Mol. Spectrosc. **63**, 120 (1976).
44. J. Liu, M.-W. Chen, D. Melnik, J. T. Yi and T. A. Miller, J. Chem. Phys. **130**, 074302 (2009).
45. F. S. Ham, Phys. Rev. **138**, A1727 (1965).
46. J. K. G. Watson, Mol. Phys. **15**, 479 (1968).
47. J. H. Van Vleck, Rev. Mod. Phys. **23**, 213 (1951).
48. J. M. Brown and T. J. Sears, J. Mol. Spectrosc. **75**, 111 (1979).
49. R. N. Zare, *Angular Momentum*. (Wiley, New York, 1988)
50. J. Liu, Ph.D. Thesis, The Ohio State University, 2007.
51. A. Rubinowicz, Z. Physik **61**, 338 (1930).
52. A. Rubinowicz, Z. Physik **65**, 662 (1930).
53. E. U. Condon and G. H. Shortley, *The Theory of Atomic Spectra*. (Cambridge University Press, London, 1935) p. 95, 100, 252-253.
54. R. F. Curl and J. L. Kinsey, J. Chem. Phys. **35**, 1758 (1961).
55. W. T. Raynes, J. Chem. Phys. **41**, 3020 (1964).
56. J. K. G. Watson, J. Mol. Spectrosc. **103**, 125 (1984).
57. J. Liu, M.-W. Chen, D. Melnik, T. A. Miller, Y. Endo and E. Hirota, J. Chem. Phys. **130**, 074303 (2009).
58. J. T. Hougen, J. Chem. Phys. **38**, 1167 (1963).
59. T. Oka, J. Chem. Phys. **47**, 5410 (1967).
60. X. Liu, L. Yu and T. A. Miller, J. Mol. Spectrosc. **140**, 112 (1990).
61. X. Liu and T. A. Miller, Mol. Phys. **75**, 1237 (1992).
62. X. Liu, S. C. Foster, J. M. Williamson, L. Yu and T. A. Miller, Mol. Phys. **69**, 357 (1990).
63. G. R. Bird, J. C. Baird, A. W. Jache, J. A. Hodgeson, R. F. Curl, A. C. Kunkle, J. W. Bransford, J. Rastrup‐Andersen and J. Rosenthal, J. Chem. Phys. **40**, 3378 (1964).
64. R. F. Curl, J. Chem. Phys. **37**, 779 (1962).
65. R. F. Curl, Mol. Phys. **9**, 585 (1965).
66. W. H. Flygare, Chem. Rev. **74**, 653 (1974).
67. G. Tarczay, S. Gopalakrishnan and T. A. Miller, J. Mol. Spectrosc. **220**, 276 (2003).
68. J. M. Brown, T. J. Sears and J. K. G. Watson, Mol. Phys. **41**, 173 (1980).
69. C. Puzzarini, J. F. Stanton and J. Gauss, Int. Rev. Phys. Chem. **29**, 273 (2010).
70. J. M. Brown and J. K. G. Watson, J. Mol. Spectrosc. **65**, 65 (1977).
71. S. Gopalakrishnan, C. C. Carter, L. Zu, V. Stakhursky, G. Tarczay and T. A. Miller, J. Chem. Phys. **118**, 4954 (2003).
72. G. M. P. Just, P. Rupper, T. A. Miller and W. L. Meerts, J. Chem. Phys. **131**, 184303 (2009).
73. J. Brown and A. Carrington, *Rotational Spectroscopy of Diatomic Molecules*. (Cambridge University Press, Cambridge, 2003)
74. V. Stakhursky and T. A. Miller, in *International Symposium on Molecular Spectroscopy* (Columbus, OH, 2001).
75. L. Cheng, (Private communications).
76. H. J. Wörner, R. van der Veen and F. Merkt, Phys. Rev. Lett. **97**, 173003 (2006).
77. H. J. Wörner and F. Merkt, J. Chem. Phys. **127**, 034303 (2007).
78. H. J. Wörner and F. Merkt, Angew. Chem. Int. Ed. **48**, 6404 (2009).
79. C. R. Brazier, L. C. Ellingboe, S. Kinsey-Nielsen and P. F. Bernath, J. Am. Chem. Soc. **108**, 2126 (1986).

80. C. R. Brazier and P. F. Bernath, J. Chem. Phys. **86**, 5918 (1987).
81. C. R. Brazier and P. F. Bernath, J. Chem. Phys. **91**, 4548 (1989).
82. P. F. Bernath, Science **254**, 665 (1991).
83. A. J. Marr, F. Grieman and T. C. Steimle, J. Chem. Phys. **105**, 3930 (1996).
84. K.-i. C. Namiki, J. S. Robinson and T. C. Steimle, J. Chem. Phys. **109**, 5283 (1998).
85. P. Crozet, F. Martin, A. J. Ross, C. Linton, M. Dick and A. G. Adam, J. Mol. Spectrosc. **213**, 28 (2002).
86. A. C. Paul, M. A. Reza and J. Liu, J. Mol. Spectrosc. **330**, 142 (2016).
87. T. A. Isaev and R. Berger, Phys. Rev. Lett. **116**, 063006 (2016).
88. I. Kozyryev, L. Baum, K. Matsuda and J. M. Doyle, ChemPhysChem **17**, 3641 (2016).
89. I. Kozyryev and N. R. Hutzler, Phys. Rev. Lett. **119**, 133002 (2017).
90. J. Jin, I. Sioutis, G. Tarczay, S. Gopalakrishnan, A. Bezant and T. A. Miller, J. Chem. Phys. **121**, 11780 (2004).

Table 1. Molecular constants of the ground-state T and G conformers of 1-propoxy. Fit 1 uses the isolated-states Hamiltonian,[71] whereas Fit 2 and Fit 3 use the coupled-states Hamiltonian proposed in the present work.

| conformation | | trans | | | | gauche | | | |
|---|---|---|---|---|---|---|---|---|---|
| symmetry | | $C_s$ | | | | $C_1$ | | | |
| | | Fit 1 [a] | Fit 2 [b] | Fit 3 [b] | calc. [c] | Fit 1 [a] | Fit 2 [b] | Fit 3 [b] | calc. |
| $\Delta E^{\tilde{A}-\tilde{X}}$ | (cm$^{-1}$) | - | 321 [d] | 321 [d] | - | - | 214 [d] | 214 [d] | - |
| $\Delta E_0$ | (cm$^{-1}$) | - | 320(5) [e] | 319.4(2) [e] | - | - | 213(3) [e] | 212.98(6) [e] | - |
| $a\zeta_e d$ | (cm$^{-1}$) | - | -25(5) | -32.3(2) | - | - | -15(3) | -20.85(6) | - |
| $A$ | (GHz) | 27.46(3) | 27.88(13) | 27.74(2) | 27.875 | 15.430(6) | 15.480(3) | 15.445(3) | 15.189 |
| $B$ | (GHz) | 3.955(4) | 3.959(4) | 3.959(4) | 3.897 | 5.211(3) | 5.213(7) | 5.204(2) | 5.221 |
| $C$ | (GHz) | 3.702(2) | 3.701(3) | 3.701(3) | 3.649 | 4.433(2) | 4.431(1) | 4.431(1) | 4.401 |
| $\zeta_t$ | | - | 1.2(2) | 0.942 [f] | | - | 1.3(2) | 0.942 [f] | |
| $\varepsilon_{aa}$ | (GHz) | -8.72(3) | 0 [f] | 0 [f] | -25.892 | -1.58(1) | 0 [f] | 0 [f] | -5.229 |
| $\varepsilon_{bb}$ | (GHz) | -0.22(2) | 0 [f] | 0 [f] | -1.012 | -1.07(2) | 0 [f] | 0 [f] | -3.485 |
| $\varepsilon_{cc}$ | (GHz) | -0.03(3) | 0 [f] | 0 [f] | 0.015 | -0.18(1) | 0 [f] | 0 [f] | -0.425 |
| $|(\varepsilon_{ab}+\varepsilon_{ba})/2|$ | (GHz) | 0.81(4) | 0 [f] | 0 [f] | 0.551 | 1.24(5) | 0 [f] | 0 [f] | 0.377 |
| $|(\varepsilon_{bc}+\varepsilon_{cb})/2|$ | (GHz) | 0 [f] | 0 [f] | 0 [f] | | 0.36(5) | 0 [f] | 0 [f] | -0.122 |
| $|(\varepsilon_{ca}+\varepsilon_{ac})/2|$ | (GHz) | 0 [f] | 0 [f] | 0 [f] | | 0.4(1) | 0 [f] | 0 [f] | 0.111 |
| $\theta$ | (deg.) | - | 24.4(4) | 24.7(3) | 24.0 | - | 57.5(1) | 57.4(1) | 56.6 |
| $\phi$ | (deg.) | - | 0 [f] | 0 [f] | 0 | - | 24.5(6) | 25.6(3) | 18.4 |
| Number of transitions | | 79 | 79 | 79 | - | 140 | 140 | 140 | - |
| rms of residules | (MHz) | 124 | 118 | 121 | - | 76 | 63 | 64 | - |

a. Ref. [71].
b. This work.
c. Calculated at the B3LYP/6-31+G** level of theory.
d. Fixed to experimentally determined $\tilde{A}-\tilde{X}$ separation.[90]
e. Calculated using $\Delta E_0 = \sqrt{\left(\Delta E^{\tilde{A}-\tilde{X}}\right)^2 - \left(a\zeta_e d\right)^2}$.
f. Fixed.
g. Numbers in parentheses are one standard deviation in the last digit.

**Table of Figures:**

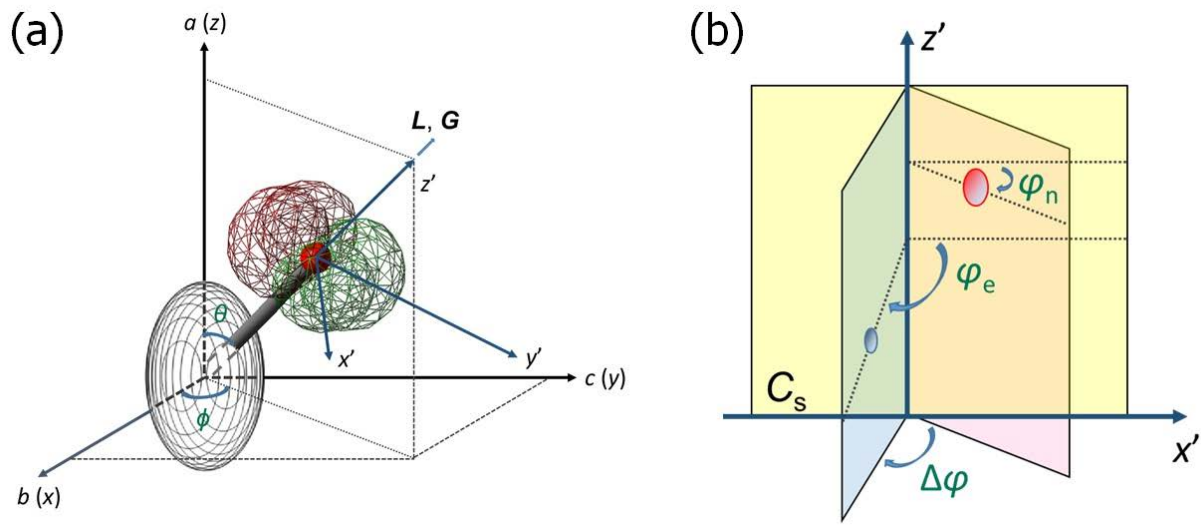

Figure 1. (a) Principal axis system (PAS) ($x$, $y$, $z$) and internal axis system (IAS) $(x', y', z')$. $a$, $b$, and $c$ are principal axes of the moment of inertia ellipsoid. Principal axes and molecule-fixed coordinates are associated in the $I^r$ representation so that $z = a$, $x = b$, $y = c$. See text for definitions of $x'$, $y'$, and $z'$. (b) Definitions of $\varphi_e$, $\varphi_n$, and $\Delta\varphi = \varphi_e - \varphi_n$. The blue and red spheres represent the unpaired electron and the atom on which it is localized, respectively.

Figure 2. Structure of Hamiltonian matrix in different symmetrized basis sets for given $J$ and $S$ values. Using case (a) basis sets, each block has a dimension of $(2J+1)$ for $P=-J,\ -J+1\ldots J-1,\ J$. Using case (b) basis sets, each block has a dimension of $(2N+1)$ for $K=-N,\ -N+1\ldots N-1,\ N$. Superscript $\alpha$ ($\alpha=x,y,z$) denotes Hamiltonian terms associated with $L_{\alpha'}$ (for $H_{SO}$), $L_{\alpha'}+G_{\alpha'}$ (for $H_C$), or $N_\alpha^2$ (for $H_r$). Note that $H_{JT}$ is not included (see text for details).

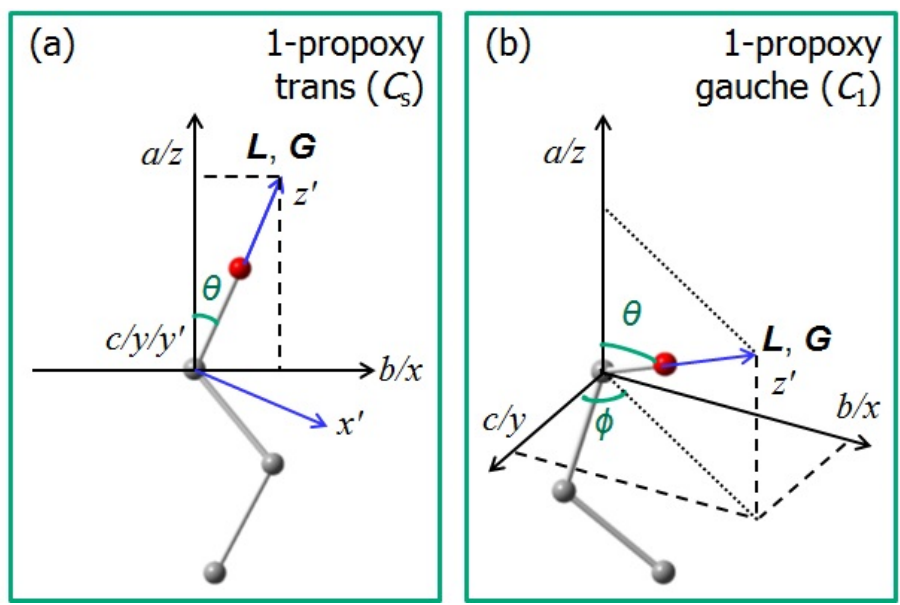

Figure 3. The two conformers of 1-propoxy in the internal $(x',y',z')$ and principal $(x,y,z)$ axis systems. The $x'$ (within the OCC plane) and the $y'$ (perpendicular to the OC$_\alpha$C$_\beta$ plane) axes for the G-conformer are not shown for clarity.

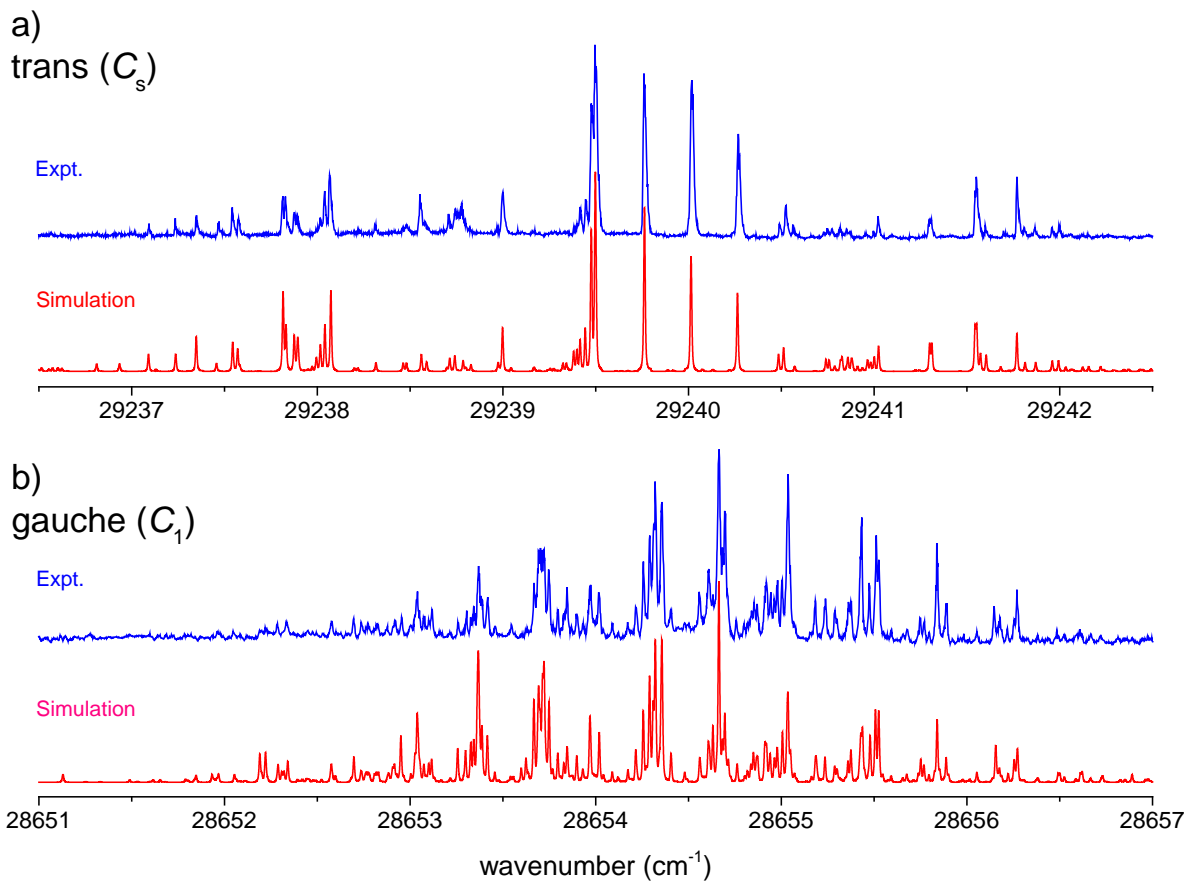

Figure 4. Experimental (blue) and simulated (red) spectra of the T- and G-conformers of the 1-propoxy radical. Molecular constants used in simulating the spectra are listed in Table 1. Weights of $a$-, $b$-, $c$-type transitions are 0:0:1 for T-conformer and 2:1:3 for G-conformer. The rotational temperature is determined to be 2 $K$ from intensity simulation. A Voigt lineshape is used to simulate the spectra. The associated Gaussian and Lorentzian profiles both have a full-width-at-half maximum (FWHM) of 75 MHz (0.0025 cm$^{-1}$).

**Supplementary Materials**

# Rotational and fine structure of open-shell molecules in nearly degenerate electronic states

Jinjun Liu

Department of Chemistry, University of Louisville, 2320 S. Brook St., Louisville, KY 40292, U.S.A.

Email: j.liu@louisville.edu

**S.1. Action rules of Hougen's operators:**

On the $|\Lambda = \pm 1\rangle$ basis set:

$$\mathcal{L}_+^2 |\Lambda = +1\rangle = 0 \qquad \mathcal{L}_+^2 |\Lambda = -1\rangle = |\Lambda = +1\rangle$$
$$\mathcal{L}_-^2 |\Lambda = +1\rangle = |\Lambda = -1\rangle \qquad \mathcal{L}_-^2 |\Lambda = -1\rangle = 0$$
$$\mathcal{L}_z |\Lambda\rangle = \Lambda |\Lambda\rangle$$

On the orbitally symmetrized case (a) basis set:

$$\left(\mathcal{L}_+^2 + \mathcal{L}_-^2\right)|\Gamma, J, P, S, \Sigma\rangle = \Gamma |\Gamma, J, P, S, \Sigma\rangle$$
$$\mathcal{L}_z |\Gamma, J, P, S, \Sigma\rangle = |-\Gamma, J, P, S, \Sigma\rangle$$

On the orbitally symmetrized case (b) basis set:

$$\left(\mathcal{L}_+^2 + \mathcal{L}_-^2\right)|\Gamma, J, N, K, S\rangle = \Gamma |\Gamma, J, N, K, S\rangle$$
$$\mathcal{L}_z |\Gamma, J, N, K, S\rangle = |-\Gamma, J, N, K, S\rangle$$

On the fully symmetrized case (a) basis set:

$$\left(\mathcal{L}_+^2 + \mathcal{L}_-^2\right)|J, \bar{P}, S, \bar{\Sigma}, \wp\rangle = \wp(-1)^{J-\bar{P}+S-\bar{\Sigma}} |J, -\bar{P}, S, -\bar{\Sigma}, \wp\rangle$$
$$\mathcal{L}_z |J, \bar{P}, S, \bar{\Sigma}, \wp\rangle = |J, \bar{P}, S, \bar{\Sigma}, -\wp\rangle$$

On the fully symmetrized case (b) basis set:

$$\left(\mathcal{L}_+^2 + \mathcal{L}_-^2\right)|J, N, \bar{K}, S, \wp\rangle = \wp(-1)^{N-\bar{K}} |J, N, -\bar{K}, S, \wp\rangle$$
$$\mathcal{L}_z |J, N, \bar{K}, S, \wp\rangle = |J, N, \bar{K}, S, -\wp\rangle$$

## S.2. Action rules of $\mathscr{L}_z$ and angular momentum operators on the orbitally symmetrized case (a) basis set:

$$\mathscr{L}_z|\Gamma,J,P,S,\Sigma\rangle = |-\Gamma,J,P,S,\Sigma\rangle$$

$$J_z|\Gamma,J,P,S,\Sigma\rangle = P|\Gamma,J,P,S,\Sigma\rangle$$

$$J_x|\Gamma,J,P,S,\Sigma\rangle = \tfrac{1}{2}\left[f(J,P)|\Gamma,J,P+1,S,\Sigma\rangle + f(J,P-1)|\Gamma,J,P-1,S,\Sigma\rangle\right]$$

$$J_y|\Gamma,J,P,S,\Sigma\rangle = \tfrac{i}{2}\left[f(J,P)|\Gamma,J,P+1,S,\Sigma\rangle - f(J,P-1)|\Gamma,J,P-1,S,\Sigma\rangle\right]$$

$$S_z|\Gamma,J,P,S,\Sigma\rangle = \Sigma|\Gamma,J,P,S,\Sigma\rangle$$

$$S_x|\Gamma,J,P,S,\Sigma\rangle = \tfrac{1}{2}|\Gamma,J,P,S,-\Sigma\rangle$$

$$S_y|\Gamma,J,P,S,\Sigma\rangle = i\Sigma|\Gamma,J,P,S,-\Sigma\rangle$$

where $i = \sqrt{-1}$ and $f(x,y) = \left[x(x+1) - y(y+1)\right]^{1/2} = \left[(x-y)(x+y+1)\right]^{1/2}$.

Note that $f(x,-y) = f(x,y-1)$ and $f(x,-y+1) = f(x,y-2)$.

## S.3. Action rules of $\mathscr{L}_z$ and angular momentum operators on the fully symmetrized case (a) basis set:

$$\mathscr{L}_z|J,\bar{P},S,\bar{\Sigma},\wp\rangle = |J,\bar{P},S,\bar{\Sigma},-\wp\rangle$$

$$J_z|J,\bar{P},S,\bar{\Sigma},\wp\rangle = \bar{P}|J,\bar{P},S,\bar{\Sigma},-\wp\rangle$$

$$J_x|J,\bar{P},S,\bar{\Sigma},\wp\rangle = \tfrac{1}{2}\left[f(J,\bar{P})|J,\bar{P}+1,S,\bar{\Sigma},-\wp\rangle + f(J,\bar{P}-1)|J,\bar{P}-1,S,\bar{\Sigma},-\wp\rangle\right]$$

$$J_y|J,\bar{P},S,\bar{\Sigma},\wp\rangle = \tfrac{i}{2}\left[f(J,\bar{P})|J,\bar{P}+1,S,\bar{\Sigma},\wp\rangle - f(J,\bar{P}-1)|J,\bar{P}-1,S,\bar{\Sigma},\wp\rangle\right]$$

$$S_z|J,\bar{P},S,\bar{\Sigma},\wp\rangle = \bar{\Sigma}|J,\bar{P},S,\bar{\Sigma},-\wp\rangle$$

$$S_x|J,\bar{P},S,\bar{\Sigma},\wp\rangle = \tfrac{1}{2}|J,\bar{P},S,-\bar{\Sigma},-\wp\rangle$$

$$S_y|J,\bar{P},S,\bar{\Sigma},\wp\rangle = i\bar{\Sigma}|J,\bar{P},S,-\bar{\Sigma},\wp\rangle$$

**S.4. Nonzero matrix elements of $H_{eff}$ in the orbitally symmetrized case (a) basis set:**

$H_q$:

$$\langle \Gamma, J, P, S, \Sigma | H_q | \Gamma, J, \bar{P}, S, \bar{\Sigma} \rangle = \tfrac{1}{2}\Gamma \Delta E_0$$

$H_{SO}$:

$$\langle -\Gamma, J, P, S, \Sigma | H_{SO} | \Gamma, J, P, S, \Sigma \rangle = a\zeta_e^z d\Sigma$$
$$\langle -\Gamma, J, P, S, -\Sigma | H_{SO} | \Gamma, J, P, S, \Sigma \rangle = \tfrac{1}{2}a\zeta_e^x d$$
$$\langle -\Gamma, J, P, S, -\Sigma | H_{SO} | \Gamma, J, P, S, \Sigma \rangle = ia\zeta_e^y d\Sigma$$

where $i = \sqrt{-1}$. $\zeta_e^z = \zeta_e \cos\theta$, $\zeta_e^x = \zeta_e \sin\theta\cos\phi$, and $\zeta_e^y = \zeta_e \sin\theta\sin\phi$.

$H_C$:

$$\langle -\Gamma, J, P, S, \Sigma | H_C | \Gamma, J, P, S, \Sigma \rangle = -2B_z\zeta_t^z(P-\Sigma)$$
$$\langle -\Gamma, J, P+1, S, \Sigma | H_C | \Gamma, J, P, S, \Sigma \rangle = -B_x\zeta_t^x f(J,P)$$
$$\langle -\Gamma, J, P-1, S, \Sigma | H_C | \Gamma, J, P, S, \Sigma \rangle = -B_x\zeta_t^x f(J,P-1)$$
$$\langle -\Gamma, J, P, S, -\Sigma | H_C | \Gamma, J, P, S, \Sigma \rangle = B_x\zeta_t^x$$
$$\langle -\Gamma, J, P+1, S, \Sigma | H_C | \Gamma, J, P, S, \Sigma \rangle = -iB_y\zeta_t^y f(J,P)$$
$$\langle -\Gamma, J, P-1, S, \Sigma | H_C | \Gamma, J, P, S, \Sigma \rangle = iB_y\zeta_t^y f(J,P-1)$$
$$\langle -\Gamma, J, P, S, -\Sigma | H_C | \Gamma, J, P, S, \Sigma \rangle = i2B_y\zeta_t^y \Sigma$$

where $f(x,y) = [x(x+1) - y(y+1)]^{1/2} = [(x-y)(x+y+1)]^{1/2}$. Note that $f(x,-y) = f(x,y-1)$ and $f(x,-y+1) = f(x,y-2)$. $\zeta_t^z = \zeta_t \cos\theta$, $\zeta_t^x = \zeta_t \sin\theta\cos\phi$, and $\zeta_t^y = \zeta_t \sin\theta\sin\phi$.

$H_r$:[a]

$$\langle \Gamma,J,P,S,\Sigma|H_r|\Gamma,J,P,S,\Sigma\rangle = B_z(P-\Sigma)^2 + \tfrac{1}{2}(B_x+B_y)[J(J+1)-P^2+\tfrac{1}{2}]$$

$$\langle \Gamma,J,P+2,S,\Sigma|H_r|\Gamma,J,P,S,\Sigma\rangle = \tfrac{1}{4}(B_x-B_y)f(J,P)f(J,P+1)$$

$$\langle \Gamma,J,P-2,S,\Sigma|H_r|\Gamma,J,P,S,\Sigma\rangle = \tfrac{1}{4}(B_x-B_y)f(J,P-1)f(J,P-2)$$

$$\langle \Gamma,J,P-1,S,\Sigma=-\tfrac{1}{2}|H_r|\Gamma,J,P,S,\Sigma=+\tfrac{1}{2}\rangle = -\tfrac{1}{2}(B_x+B_y)f(J,P-1)$$

$$\langle \Gamma,J,P+1,S,\Sigma=+\tfrac{1}{2}|H_r|\Gamma,J,P,S,\Sigma=-\tfrac{1}{2}\rangle = -\tfrac{1}{2}(B_x+B_y)f(J,P)$$

$$\langle \Gamma,J,P+1,S,\Sigma=-\tfrac{1}{2}|H_r|\Gamma,J,P,S,\Sigma=+\tfrac{1}{2}\rangle = -\tfrac{1}{2}(B_x-B_y)f(J,P)$$

$$\langle \Gamma,J,P-1,S,\Sigma=+\tfrac{1}{2}|H_r|\Gamma,J,P,S,\Sigma=-\tfrac{1}{2}\rangle = -\tfrac{1}{2}(B_x-B_y)f(J,P-1)$$

$H_{SR}$:[b]

$$\langle \Gamma,J,P,S,\Sigma|H_{SR}|\Gamma,J,P,S,\Sigma\rangle = \varepsilon_{zz}(P-\Sigma)\Sigma - (\varepsilon_{xx}+\varepsilon_{yy})/4$$

$$\langle \Gamma,J,P-1,S,\Sigma=-\tfrac{1}{2}|H_{SR}|\Gamma,J,P,S,\Sigma=+\tfrac{1}{2}\rangle = \tfrac{1}{4}(\varepsilon_{xx}+\varepsilon_{yy})f(J,P-1)$$

$$\langle \Gamma,J,P+1,S,\Sigma=+\tfrac{1}{2}|H_{SR}|\Gamma,J,P,S,\Sigma=-\tfrac{1}{2}\rangle = \tfrac{1}{4}(\varepsilon_{xx}+\varepsilon_{yy})f(J,P)$$

$$\langle \Gamma,J,P+1,S,\Sigma=-\tfrac{1}{2}|H_{SR}|\Gamma,J,P,S,\Sigma=+\tfrac{1}{2}\rangle = \tfrac{1}{4}(\varepsilon_{xx}-\varepsilon_{yy})f(J,P)$$

$$\langle \Gamma,J,P-1,S,\Sigma=+\tfrac{1}{2}|H_{SR}|\Gamma,J,P,S,\Sigma=-\tfrac{1}{2}\rangle = \tfrac{1}{4}(\varepsilon_{xx}-\varepsilon_{yy})f(J,P-1)$$

$$\langle \Gamma,J,P,S,-\Sigma|H_{SR}|\Gamma,J,P,S,\Sigma\rangle = \tfrac{1}{4}(\varepsilon_{zx}+\varepsilon_{xz})P$$

$$\langle \Gamma,J,P+1,S,\Sigma|H_{SR}|\Gamma,J,P,S,\Sigma\rangle = \tfrac{1}{4}(\varepsilon_{zx}+\varepsilon_{xz})f(J,P)\Sigma$$

$$\langle \Gamma,J,P-1,S,\Sigma|H_{SR}|\Gamma,J,P,S,\Sigma\rangle = \tfrac{1}{4}(\varepsilon_{zx}+\varepsilon_{xz})f(J,P-1)\Sigma$$

$$\langle \Gamma,J,P+1,S,-\Sigma|H_{SR}|\Gamma,J,P,S,\Sigma\rangle = \tfrac{i}{4}(\Sigma+\tfrac{1}{2})f(J,P)$$

$$\langle \Gamma,J,P-1,S,-\Sigma|H_{SR}|\Gamma,J,P,S,\Sigma\rangle = \tfrac{i}{4}(\Sigma-\tfrac{1}{2})f(J,P-1)$$

$$\langle \Gamma,J,P,S,-\Sigma|H_{SR}|\Gamma,J,P,S,\Sigma\rangle = \tfrac{i}{2}(\varepsilon_{yz}+\varepsilon_{zy})P\Sigma$$

$$\langle \Gamma,J,P+1,S,\Sigma|H_{SR}|\Gamma,J,P,S,\Sigma\rangle = \tfrac{i}{4}(\varepsilon_{yz}+\varepsilon_{zy})f(J,P)\Sigma$$

$$\langle \Gamma,J,P-1,S,\Sigma|H_{SR}|\Gamma,J,P,S,\Sigma\rangle = -\tfrac{i}{4}(\varepsilon_{yz}+\varepsilon_{zy})f(J,P-1)\Sigma$$

---

[a] $H_r$ contains terms that are proportional to $J_\alpha^2$, $S_\alpha^2$, and $-2J_\alpha S_\alpha$ (the spin-uncoupling terms).

[b] $H_{SR}$ contains terms that are proportional to $J_\alpha S_\beta$ and $S_\alpha S_\beta$.

**S.5. Nonzero matrix elements of $H_{eff}$ in the fully symmetrized case (a) basis set:**

$H_q$:

$$\langle J,-\bar{P},S,-\bar{\Sigma},\wp|H_q|J,\bar{P},S,\bar{\Sigma},\wp\rangle = \tfrac{1}{2}\wp(-1)^{J-\bar{P}+S-\bar{\Sigma}}\Delta E_0$$

$H_{SO}$:

$$\langle J,\bar{P},S,\bar{\Sigma},\wp|H_{SO}|J,\bar{P},S,\Sigma,\wp\rangle = a\zeta_e^z d\bar{\Sigma}$$
$$\langle J,\bar{P},S,-\bar{\Sigma},\wp|H_{SO}|J,\bar{P},S,\Sigma,\wp\rangle = \tfrac{1}{2}a\zeta_e^x d$$
$$\langle J,\bar{P},S,-\bar{\Sigma},-\wp|H_{SO}|J,\bar{P},S,\Sigma,\wp\rangle = ia\zeta_e^y d\bar{\Sigma}$$

where $\zeta_e^z = \zeta_e\cos\theta$, $\zeta_e^x = \zeta_e\sin\theta\cos\phi$, and $\zeta_e^y = \zeta_e\sin\theta\sin\phi$.

$H_C$:

$$\langle J,\bar{P},S,\bar{\Sigma},\wp|H_C|J,\bar{P},S,\bar{\Sigma},\wp\rangle = -2B_z\zeta_t^z(\bar{P}-\bar{\Sigma})$$
$$\langle J,\bar{P}+1,S,\bar{\Sigma},\wp|H_C|J,\bar{P},S,\bar{\Sigma},\wp\rangle = -B_x\zeta_t^x f(J,\bar{P})$$
$$\langle J,\bar{P}-1,S,\bar{\Sigma},\wp|H_C|J,\bar{P},S,\bar{\Sigma},\wp\rangle = -B_x\zeta_t^x f(J,\bar{P}-1)$$
$$\langle J,\bar{P},S,-\bar{\Sigma},\wp|H_C|J,\bar{P},S,\bar{\Sigma},\wp\rangle = B_x\zeta_t^x$$
$$\langle J,\bar{P}+1,S,\bar{\Sigma},-\wp|H_C|J,\bar{P},S,\bar{\Sigma},\wp\rangle = -iB_y\zeta_t^y f(J,\bar{P})$$
$$\langle J,\bar{P}-1,S,\bar{\Sigma},-\wp|H_C|J,\bar{P},S,\bar{\Sigma},\wp\rangle = iB_y\zeta_t^y f(J,\bar{P}-1)$$
$$\langle J,\bar{P},S,-\bar{\Sigma},-\wp|H_C|J,\bar{P},S,\bar{\Sigma},\wp\rangle = i2B_y\zeta_t^y\bar{\Sigma}$$

where $\zeta_t^z = \zeta_t\cos\theta$, $\zeta_t^x = \zeta_t\sin\theta\cos\phi$, and $\zeta_t^y = \zeta_t\sin\theta\sin\phi$.

$H_r$:

$\langle J, \overline{P}, S, \overline{\Sigma}, \wp | H_r | J, \overline{P}, S, \overline{\Sigma}, \wp \rangle = B_z(\overline{P}-\overline{\Sigma})^2 + \tfrac{1}{2}(B_x+B_y)[J(J+1) - \overline{P}^2 + \tfrac{1}{2}]$

$\langle J, \overline{P}+2, S, \overline{\Sigma}, \wp | H_r | J, \overline{P}, S, \overline{\Sigma}, \wp \rangle = \tfrac{1}{4}(B_x-B_y)f(J,\overline{P})f(J,\overline{P}+1)$

$\langle J, \overline{P}-2, S, \overline{\Sigma}, \wp | H_r | J, \overline{P}, S, \overline{\Sigma}, \wp \rangle = \tfrac{1}{4}(B_x-B_y)f(J,\overline{P}-1)f(J,\overline{P}-2)$

$\langle J, \overline{P}-1, S, \overline{\Sigma}=-\tfrac{1}{2}, \wp | H_r | J, \overline{P}, S, \overline{\Sigma}=+\tfrac{1}{2}, \wp \rangle = -\tfrac{1}{2}(B_x+B_y)f(J,\overline{P}-1)$

$\langle J, \overline{P}+1, S, \overline{\Sigma}=+\tfrac{1}{2}, \wp | H_r | J, \overline{P}, S, \overline{\Sigma}=-\tfrac{1}{2}, \wp \rangle = -\tfrac{1}{2}(B_x+B_y)f(J,\overline{P})$

$\langle J, \overline{P}+1, S, \overline{\Sigma}=-\tfrac{1}{2}, \wp | H_r | J, \overline{P}, S, \overline{\Sigma}=+\tfrac{1}{2}, \wp \rangle = -\tfrac{1}{2}(B_x-B_y)f(J,\overline{P})$

$\langle J, \overline{P}-1, S, \overline{\Sigma}=+\tfrac{1}{2}, \wp | H_r | J, \overline{P}, S, \overline{\Sigma}=-\tfrac{1}{2}, \wp \rangle = -\tfrac{1}{2}(B_x-B_y)f(J,\overline{P}-1)$

$H_{SR}$:

$\langle J, \overline{P}, S, \overline{\Sigma}, \wp | H_{SR} | J, \overline{P}, S, \overline{\Sigma}, \wp \rangle = \varepsilon_{zz}(\overline{P}-\overline{\Sigma})\overline{\Sigma} - (\varepsilon_{xx}+\varepsilon_{yy})/4$

$\langle J, \overline{P}-1, S, \overline{\Sigma}=-\tfrac{1}{2}, \wp | H_{SR} | J, \overline{P}, S, \overline{\Sigma}=+\tfrac{1}{2}, \wp \rangle = \tfrac{1}{4}(\varepsilon_{xx}+\varepsilon_{yy})f(J,\overline{P}-1)$

$\langle J, \overline{P}+1, S, \overline{\Sigma}=+\tfrac{1}{2}, \wp | H_{SR} | J, \overline{P}, S, \overline{\Sigma}=-\tfrac{1}{2}, \wp \rangle = \tfrac{1}{4}(\varepsilon_{xx}+\varepsilon_{yy})f(J,\overline{P})$

$\langle J, \overline{P}+1, S, \overline{\Sigma}=-\tfrac{1}{2}, \wp | H_{SR} | J, \overline{P}, S, \overline{\Sigma}=+\tfrac{1}{2}, \wp \rangle = \tfrac{1}{4}(\varepsilon_{xx}-\varepsilon_{yy})f(J,\overline{P})$

$\langle J, \overline{P}-1, S, \overline{\Sigma}=+\tfrac{1}{2}, \wp | H_{SR} | J, \overline{P}, S, \overline{\Sigma}=-\tfrac{1}{2}, \wp \rangle = \tfrac{1}{4}(\varepsilon_{xx}-\varepsilon_{yy})f(J,\overline{P}-1)$

$\langle J, \overline{P}, S, -\overline{\Sigma}, \wp | H_{SR} | J, \overline{P}, S, \overline{\Sigma}, \wp \rangle = \tfrac{1}{4}(\varepsilon_{zx}+\varepsilon_{xz})\overline{P}$

$\langle J, \overline{P}+1, S, \overline{\Sigma}, \wp | H_{SR} | J, \overline{P}, S, \overline{\Sigma}, \wp \rangle = \tfrac{1}{4}(\varepsilon_{zx}+\varepsilon_{xz})f(J,\overline{P})\overline{\Sigma}$

$\langle J, \overline{P}-1, S, \overline{\Sigma}, \wp | H_{SR} | J, \overline{P}, S, \overline{\Sigma}, \wp \rangle = \tfrac{1}{4}(\varepsilon_{zx}+\varepsilon_{xz})f(J,\overline{P}-1)\overline{\Sigma}$

$\langle J, \overline{P}+1, S, -\overline{\Sigma}, -\wp | H_{SR} | J, \overline{P}, S, \overline{\Sigma}, \wp \rangle = \tfrac{i}{4}\left(\overline{\Sigma}+\tfrac{1}{2}\right)(\varepsilon_{xy}+\varepsilon_{yx})f(J,\overline{P})$

$\langle J, \overline{P}-1, S, -\overline{\Sigma}, -\wp | H_{SR} | J, \overline{P}, S, \overline{\Sigma}, \wp \rangle = \tfrac{i}{4}\left(\overline{\Sigma}-\tfrac{1}{2}\right)(\varepsilon_{xy}+\varepsilon_{yx})f(J,\overline{P}-1)\overline{\Sigma}$

$\langle J, \overline{P}, S, -\overline{\Sigma}, -\wp | H_{SR} | J, \overline{P}, S, \overline{\Sigma}, \wp \rangle = \tfrac{i}{2}(\varepsilon_{yz}+\varepsilon_{zy})\overline{P}\overline{\Sigma}$

$\langle J, \overline{P}+1, S, \overline{\Sigma}, -\wp | H_{SR} | J, \overline{P}, S, \overline{\Sigma}, \wp \rangle = \tfrac{i}{4}(\varepsilon_{yz}+\varepsilon_{zy})f(J,\overline{P})\overline{\Sigma}$

$\langle J, \overline{P}-1, S, \overline{\Sigma}, -\wp | H_{SR} | J, \overline{P}, S, \overline{\Sigma}, \wp \rangle = -\tfrac{i}{4}(\varepsilon_{yz}+\varepsilon_{zy})f(J,\overline{P}-1)\overline{\Sigma}$

**S.6. Nonzero matrix elements of $H_{eff}$ in the orbitally symmetrized case (b) basis set:**

$H_q$:

$$\langle \Gamma, J, N, K, S | H_q | \Gamma, J, N, K, S \rangle = \tfrac{1}{2}\Gamma \Delta E_0$$

$H_{so}$:

$$\langle -\Gamma, J, N, K, S | H_{so} | \Gamma, J, N, K, S \rangle = \mp a\zeta_e^z d \frac{K}{2J+1}$$

$$\langle -\Gamma, J, N \mp 1, K, S | H_{so} | \Gamma, J, N, K, S \rangle = a\zeta_e^z d \frac{\left[(J+\tfrac{1}{2})^2 - K^2\right]^{1/2}}{2J+1}$$

$$\langle -\Gamma, J, N, K+1, S | H_{so} | \Gamma, J, N, K, S \rangle = \mp \frac{a\zeta_e^x d}{2(2J+1)} f(N,K)$$

$$\langle -\Gamma, J, N, K-1, S | H_{so} | \Gamma, J, N, K, S \rangle = \mp \frac{a\zeta_e^x d}{2(2J+1)} f(N,K-1)$$

$$\langle -\Gamma, J, N \mp 1, K+1, S | H_{so} | \Gamma, J, N, K, S \rangle = \pm \frac{a\zeta_e^x d}{2(2J+1)} \left\{ (J \mp K + \tfrac{1}{2}) \left[ J \mp (K+1) + \tfrac{1}{2} \right] \right\}^{1/2}$$

$$\langle -\Gamma, J, N \mp 1, K-1, S | H_{so} | \Gamma, J, N, K, S \rangle = \mp \frac{a\zeta_e^x d}{2(2J+1)} \left\{ (J \pm K + \tfrac{1}{2}) \left[ J \pm (K-1) + \tfrac{1}{2} \right] \right\}^{1/2}$$

$$\langle -\Gamma, J, N, K+1, S | H_{so} | \Gamma, J, N, K, S \rangle = \mp \frac{ia\zeta_e^y d}{2(2J+1)} f(N,K)$$

$$\langle -\Gamma, J, N, K-1, S | H_{so} | \Gamma, J, N, K, S \rangle = \pm \frac{ia\zeta_e^y d}{2(2J+1)} f(N,K-1)$$

$$\langle -\Gamma, J, N \mp 1, K+1, S | H_{so} | \Gamma, J, N, K, S \rangle = \pm \frac{ia\zeta_e^y d}{2(2J+1)} \left\{ (J \mp K + \tfrac{1}{2}) \left[ J \mp (K+1) + \tfrac{1}{2} \right] \right\}^{1/2}$$

$$\langle -\Gamma, J, N \mp 1, K-1, S | H_{so} | \Gamma, J, N, K, S \rangle = \pm \frac{ia\zeta_e^y d}{2(2J+1)} \left\{ (J \pm K + \tfrac{1}{2}) \left[ J \pm (K-1) + \tfrac{1}{2} \right] \right\}^{1/2}$$

with "$\pm$" for $N = J \pm \tfrac{1}{2}$, respectively.

where $\zeta_e^z = \zeta_e \cos\theta$, $\zeta_e^x = \zeta_e \sin\theta \cos\phi$, and $\zeta_e^y = \zeta_e \sin\theta \sin\phi$.

$H_c$:

$\langle -\Gamma, J, N, K, S | H_c | \Gamma, J, N, K, S \rangle = -2B_z \zeta_t^z K$

$\langle -\Gamma, J, N, K+1, S | H_c | \Gamma, J, N, K, S \rangle = -B_x \zeta_t^x f(N, K)$

$\langle -\Gamma, J, N, K-1, S | H_c | \Gamma, J, N, K, S \rangle = -B_x \zeta_t^x f(N, K-1)$

$\langle -\Gamma, J, N, K+1, S | H_c | \Gamma, J, N, K, S \rangle = iB_y \zeta_t^y f(N, K)$

$\langle -\Gamma, J, N, K-1, S | H_c | \Gamma, J, N, K, S \rangle = -iB_y \zeta_t^y f(N, K-1)$

where $\zeta_t^z = \zeta_t \cos\theta$, $\zeta_t^x = \zeta_t \sin\theta \cos\phi$, and $\zeta_t^y = \zeta_t \sin\theta \sin\phi$.

$H_r$:[1,2]

$\langle \Gamma, J, N, K, S | H_r | \Gamma, J, N, K, S \rangle = \frac{1}{2}(B_x + B_y)[N(N+1) - K^2] + B_z K^2$

$\langle \Gamma, J, N, K+2, S | H_r | \Gamma, J, N, K, S \rangle = \frac{1}{2}(B_x - B_y)[\frac{1}{2} f(N, K) f(N, K+1)]$

$\langle \Gamma, J, N, K-2, S | H_r | \Gamma, J, N, K, S \rangle = \frac{1}{2}(B_x - B_y)[\frac{1}{2} f(N, K-1) f(N, K-2)]$

$H_{SR}$:[1,2]

$$\langle \Gamma, J, N, K, S | H_{SR} | \Gamma, J, N, K, S \rangle = -\tfrac{1}{2} a_0 [J(J+1) - N(N+1) - S(S+1)] + a[3K^2 - N(N+1)]\theta(N)$$

$$\langle \Gamma, J, N, K+1, S | H_{SR} | \Gamma, J, N, K, S \rangle = (d+ie)(K+\tfrac{1}{2})f(N,K)\theta(N)$$

$$\langle \Gamma, J, N, K-1, S | H_{SR} | \Gamma, J, N, K, S \rangle = (d-ie)(K-\tfrac{1}{2})f(N,K-1)\theta(N)$$

$$\langle \Gamma, J, N, K+2, S | H_{SR} | \Gamma, J, N, K, S \rangle = \tfrac{1}{2}(b+ic)f(N,K)f(N,K+1)\theta(N)$$

$$\langle \Gamma, J, N, K-2, S | H_{SR} | \Gamma, J, N, K, S \rangle = \tfrac{1}{2}(b-ic)f(N,K-1)f(N,K-2)\theta(N)$$

$$\langle \Gamma, J, N-1, K, S | H_{SR} | \Gamma, J, N, K, S \rangle = \tfrac{3}{2} aK(N^2 - K^2)^{\tfrac{1}{2}}\varphi(N) \text{ for } N = J + \tfrac{1}{2}$$

$$\langle \Gamma, J, N+1, K, S | H_{SR} | \Gamma, J, N, K, S \rangle = \tfrac{3}{2} aK\left[(N+1)^2 - K^2\right]^{\tfrac{1}{2}} \varphi(N+1) \text{ for } N = J - \tfrac{1}{2}$$

$$\langle \Gamma, J, N-1, K+1, S | H_{SR} | \Gamma, J, N, K, S \rangle = \tfrac{1}{4}(d+ie)(N+2K+1)g(N,K)\varphi(N) \text{ for } N = J + \tfrac{1}{2}$$

$$\langle \Gamma, J, N+1, K-1, S | H_{SR} | \Gamma, J, N, K, S \rangle = \tfrac{1}{4}(d-ie)(N+2K)g(N+1,K-1)\varphi(N+1) \text{ for } N = J - \tfrac{1}{2}$$

$$\langle \Gamma, J, N-1, K-1, S | H_{SR} | \Gamma, J, N, K, S \rangle = \tfrac{1}{4}(d-ie)(N-2K+1)g(N,-K)\varphi(N) \text{ for } N = J + \tfrac{1}{2}$$

$$\langle \Gamma, J, N+1, K+1, S | H_{SR} | \Gamma, J, N, K, S \rangle = \tfrac{1}{4}(d+ie)(N-2K)g(N+1,-K-1)\varphi(N+1) \text{ for } N = J - \tfrac{1}{2}$$

$$\langle \Gamma, J, N-1, K+2, S | H_{SR} | \Gamma, J, N, K, S \rangle = \tfrac{1}{4}(b+ic)f(N,K)g(N,K+1)\varphi(N) \text{ for } N = J + \tfrac{1}{2}$$

$$\langle \Gamma, J, N+1, K-2, S | H_{SR} | \Gamma, J, N, K, S \rangle = \tfrac{1}{4}(b-ic)f(N+1,K-2)g(N+1,K-1)\varphi(N+1) \text{ for } N = J - \tfrac{1}{2}$$

$$\langle \Gamma, J, N-1, K-2, S | H_{SR} | \Gamma, J, N, K, S \rangle = -\tfrac{1}{4}(b-ic)f(N,K-1)g(N,-K+1)\varphi(N) \text{ for } N = J + \tfrac{1}{2}$$

$$\langle \Gamma, J, N+1, K+2, S | H_{SR} | \Gamma, J, N, K, S \rangle = -\tfrac{1}{4}(b+ic)f(N+1,K+1)g(N+1,-K-1)\varphi(N+1) \text{ for } N = J - \tfrac{1}{2}$$

where

$g(x,y) = [(x-y)(x-y-1)]^{1/2}$,

$\theta(N) = \dfrac{-C(N)}{2N(N+1)}$, with $C(N) = J(J+1) - N(N+1) - S(S+1)$

$\varphi(N) = -\dfrac{1}{N}\left[\dfrac{P(N)Q(N-1)}{(2N-1)(2N+1)}\right]^{1/2}$ with $P(N) = (N - J + S)(N + J + S + 1)$,

and $Q(N) = (S + J - N)(N + J - S + 1)$.

The SR Hamiltonian matrix above is computed using the irreducible tensor SR constants, which are related to SR constants in the PAS as follows:

$$T_0^0(\varepsilon) = (-1/\sqrt{3})(\varepsilon_{zz} + \varepsilon_{xx} + \varepsilon_{yy}) = \sqrt{3} a_0$$

$$T_0^2(\varepsilon) = (1/\sqrt{6})(2\varepsilon_{zz} - \varepsilon_{xx} - \varepsilon_{yy}) = -\sqrt{6}\, a$$

$$T_{\pm 1}^2(\varepsilon) = \mp (1/2)[(\varepsilon_{zx} + \varepsilon_{xz}) \pm i(\varepsilon_{zy} + \varepsilon_{yz})] = \pm(d \pm ie)$$

$$T_{\pm 2}^2(\varepsilon) = (1/2)[(\varepsilon_{xx} - \varepsilon_{yy}) \pm i(\varepsilon_{xy} + \varepsilon_{yx})] = b \pm ic$$

The $T_{0,\pm 1}^1(\varepsilon)$ components vanish for symmetry reasons. [3]

**S.7. Nonzero matrix elements of $H_{eff}$ in the fully symmetrized case (b) basis set:**

$H_q$:

$$\langle J, N, -\bar{K}, S, \wp | H_q | J, N, \bar{K}, S, \wp \rangle = \tfrac{1}{2}\wp(-1)^{N-\bar{K}} \Delta E_0$$

$H_{so}$:

$$\langle J, N, \bar{K}, S, \wp | H_{so} | J, N, \bar{K}, S, \wp \rangle = \mp a\zeta_e^z d \frac{\bar{K}}{2J+1}$$

$$\langle J, N\mp 1, \bar{K}, S, \wp | H_{so} | J, N, \bar{K}, S, \wp \rangle = a\zeta_e^z d \frac{\left[(J+\tfrac{1}{2})^2 - \bar{K}^2\right]^{1/2}}{2J+1}$$

$$\langle J, N, \bar{K}+1, S, \wp | H_{so} | J, N, \bar{K}, S, \wp \rangle = \mp \frac{a\zeta_e^x d}{2(2J+1)} f(N, \bar{K})$$

$$\langle J, N, \bar{K}-1, S, \wp | H_{so} | J, N, \bar{K}, S, \wp \rangle = \mp \frac{a\zeta_e^x d}{2(2J+1)} f(N, \bar{K}-1)$$

$$\langle J, N\mp 1, \bar{K}+1, S, \wp | H_{so} | J, N, \bar{K}, S, \wp \rangle = \pm \frac{a\zeta_e^x d}{2(2J+1)} \left\{ (J\mp \bar{K}+\tfrac{1}{2})\left[J \mp (\bar{K}+1)+\tfrac{1}{2}\right]\right\}^{1/2}$$

$$\langle J, N\mp 1, \bar{K}-1, S, \wp | H_{so} | J, N, \bar{K}, S, \wp \rangle = \mp \frac{a\zeta_e^x d}{2(2J+1)} \left\{ (J\pm \bar{K}+\tfrac{1}{2})\left[J \pm (\bar{K}-1)+\tfrac{1}{2}\right]\right\}^{1/2}$$

$$\langle J, N, \bar{K}+1, S, -\wp | H_{so} | J, N, \bar{K}, S, \wp \rangle = \mp \frac{ia\zeta_e^y d}{2(2J+1)} f(N, \bar{K})$$

$$\langle J, N, \bar{K}-1, S, -\wp | H_{so} | J, N, \bar{K}, S, \wp \rangle = \pm \frac{ia\zeta_e^y d}{2(2J+1)} f(N, \bar{K}-1)$$

$$\langle J, N\mp 1, \bar{K}+1, S, -\wp | H_{so} | J, N, \bar{K}, S, \wp \rangle = \pm \frac{ia\zeta_e^y d}{2(2J+1)} \left\{ (J\mp \bar{K}+\tfrac{1}{2})\left[J \mp (\bar{K}+1)+\tfrac{1}{2}\right]\right\}^{1/2}$$

$$\langle J, N\mp 1, \bar{K}-1, S, -\wp | H_{so} | J, N, \bar{K}, S, \wp \rangle = \pm \frac{ia\zeta_e^y d}{2(2J+1)} \left\{ (J\pm \bar{K}+\tfrac{1}{2})\left[J \pm (\bar{K}-1)+\tfrac{1}{2}\right]\right\}^{1/2}$$

with "$\pm$" for $N = J \pm \tfrac{1}{2}$, respectively.

where $\zeta_e^z = \zeta_e \cos\theta$, $\zeta_e^x = \zeta_e \sin\theta \cos\phi$, and $\zeta_e^y = \zeta_e \sin\theta \sin\phi$.

$H_c$:

$\langle J, N, \bar{K}, S, \wp | H_c | J, N, \bar{K}, S, \wp \rangle = -2 B_z \zeta_t^z \bar{K}$

$\langle J, N, \bar{K}+1, S, \wp | H_c | J, N, \bar{K}, S, \wp \rangle = -B_x \zeta_t^x f(N, \bar{K})$

$\langle J, N, \bar{K}-1, S, \wp | H_c | J, N, \bar{K}, S, \wp \rangle = -B_x \zeta_t^x f(N, \bar{K}-1)$

$\langle J, N, \bar{K}+1, S, -\wp | H_c | J, N, \bar{K}, S, \wp \rangle = i B_y \zeta_t^y f(N, \bar{K})$

$\langle J, N, \bar{K}-1, S, -\wp | H_c | J, N, \bar{K}, S, \wp \rangle = -i B_y \zeta_t^y f(N, \bar{K}-1)$

where $\zeta_t^z = \zeta_t \cos\theta$, $\zeta_t^x = \zeta_t \sin\theta \cos\phi$, and $\zeta_t^y = \zeta_t \sin\theta \sin\phi$.

$H_r$:

$\langle J, N, \bar{K}, S, \wp | H_r | J, N, \bar{K}, S, \wp \rangle = \tfrac{1}{2}(B_x + B_y)[N(N+1) - \bar{K}^2] + B_z \bar{K}^2$

$\langle J, N, \bar{K}+2, S, \wp | H_r | J, N, \bar{K}, S, \wp \rangle = \tfrac{1}{2}(B_x - B_y)[\tfrac{1}{2} f(N, \bar{K}) f(N, \bar{K}+1)]$

$\langle J, N, \bar{K}-2, S, \wp | H_r | J, N, \bar{K}, S, \wp \rangle = \tfrac{1}{2}(B_x - B_y)[\tfrac{1}{2} f(N, \bar{K}-1) f(N, \bar{K}-2)]$

$H_{SR}$:

$\langle N, \bar{K}, S, \wp | H_{SR} | J, N, \bar{K}, S, \wp \rangle = -\tfrac{1}{2} a_0 [J(J+1) - N(N+1) - S(S+1)] + a[3\bar{K}^2 - N(N+1)]\theta(N)$

$\langle J, N, \bar{K}+1, S, \wp | H_{SR} | J, N, \bar{K}, S, \wp \rangle = d(\bar{K} + \tfrac{1}{2}) f(N, \bar{K}) \theta(N)$

$\langle J, N, \bar{K}+1, S, -\wp | H_{SR} | J, N, \bar{K}, S, \wp \rangle = ie(\bar{K} + \tfrac{1}{2}) f(N, \bar{K}) \theta(N)$

$\langle J, N, \bar{K}-1, S, \wp | H_{SR} | J, N, \bar{K}, S, \wp \rangle = d(\bar{K} - \tfrac{1}{2}) f(N, \bar{K}-1) \theta(N)$

$\langle J, N, \bar{K}-1, S, -\wp | H_{SR} | J, N, \bar{K}, S, \wp \rangle = -ie(\bar{K} - \tfrac{1}{2}) f(N, \bar{K}-1) \theta(N)$

$\langle J, N, \bar{K}+2, S, \wp | H_{SR} | J, N, \bar{K}, S, \wp \rangle = \tfrac{1}{2} b f(N, \bar{K}) f(N, \bar{K}+1) \theta(N)$

$\langle J, N, \bar{K}+2, S, -\wp | H_{SR} | J, N, \bar{K}, S, \wp \rangle = \tfrac{i}{2} c f(N, \bar{K}) f(N, \bar{K}+1) \theta(N)$

$\langle J, N, \bar{K}-2, S, \wp | H_{SR} | J, N, \bar{K}, S, \wp \rangle = \tfrac{1}{2} b f(N, \bar{K}-1) f(N, \bar{K}-2) \theta(N)$

$\langle J, N, \bar{K}-2, S, -\wp | H_{SR} | J, N, \bar{K}, S, \wp \rangle = -\tfrac{i}{2} c f(N, \bar{K}-1) f(N, \bar{K}-2) \theta(N)$

$\langle J, N-1, \bar{K}, S, \wp | H_{SR} | J, N, \bar{K}, S, \wp \rangle = \tfrac{3}{2} a \bar{K} (N^2 - \bar{K}^2)^{\tfrac{1}{2}} \varphi(N)$ for $N = J + \tfrac{1}{2}$

$\langle J, N-1, \bar{K}, S, \wp | H_{SR} | J, N, \bar{K}, S, \wp \rangle = \tfrac{3}{2} a \bar{K} \left[ (N+1)^2 - \bar{K}^2 \right]^{\tfrac{1}{2}} \varphi(N+1)$ for $N = J - \tfrac{1}{2}$

$\langle J, N-1, \bar{K}+1, S, \wp | H_{SR} | J, N, \bar{K}, S, \wp \rangle = \tfrac{1}{4} d(N + 2\bar{K} + 1) g(N, \bar{K}) \varphi(N)$ for $N = J + \tfrac{1}{2}$

$\langle J, N-1, \bar{K}+1, S, -\wp | H_{SR} | J, N, \bar{K}, S, \wp \rangle = \tfrac{i}{4} e(N + 2\bar{K} + 1) g(N, \bar{K}) \varphi(N)$ for $N = J + \tfrac{1}{2}$

$\langle J, N-1, \bar{K}+1, S, \wp | H_{SR} | J, N, \bar{K}, S, \wp \rangle = \tfrac{1}{4} d(N + 2\bar{K}) g(N+1, \bar{K}-1) \varphi(N+1)$ for $N = J - \tfrac{1}{2}$

$\langle J, N-1, \bar{K}+1, S, -\wp | H_{SR} | J, N, \bar{K}, S, \wp \rangle = \tfrac{i}{4} e(N + 2\bar{K}) g(N+1, \bar{K}-1) \varphi(N+1)$ for $N = J - \tfrac{1}{2}$

$\langle J, N-1, \bar{K}-1, S, \wp | H_{SR} | J, N, \bar{K}, S, \wp \rangle = \tfrac{1}{4} d(N - 2\bar{K} + 1) g(N, -\bar{K}) \varphi(N)$ for $N = J + \tfrac{1}{2}$

$\langle J, N-1, \bar{K}-1, S, -\wp | H_{SR} | J, N, \bar{K}, S, \wp \rangle = -\tfrac{i}{4} e(N - 2\bar{K} + 1) g(N, -\bar{K}) \varphi(N)$ for $N = J + \tfrac{1}{2}$

$\langle J, N-1, \bar{K}-1, S, \wp | H_{SR} | J, N, \bar{K}, S, \wp \rangle = \tfrac{1}{4} d(N - 2\bar{K}) g(N+1, -\bar{K}-1) \varphi(N+1)$ for $N = J - \tfrac{1}{2}$

$\langle J, N-1, \bar{K}-1, S, -\wp | H_{SR} | J, N, \bar{K}, S, \wp \rangle = -\tfrac{i}{4} e(N - 2\bar{K}) g(N+1, -\bar{K}-1) \varphi(N+1)$ for $N = J - \tfrac{1}{2}$

$\langle J, N-1, \bar{K}+2, S, \wp | H_{SR} | J, N, \bar{K}, S, \wp \rangle = \tfrac{1}{4} b f(N, \bar{K}) g(N, \bar{K}+1) \varphi(N)$ for $N = J + \tfrac{1}{2}$

$\langle J, N-1, \bar{K}+2, S, -\wp | H_{SR} | J, N, \bar{K}, S, \wp \rangle = \tfrac{i}{4} c f(N, \bar{K}) g(N, \bar{K}+1) \varphi(N)$ for $N = J + \tfrac{1}{2}$

$\langle J, N-1, \bar{K}+2, S, \wp | H_{SR} | J, N, \bar{K}, S, \wp \rangle = \tfrac{1}{4} b f(N+1, \bar{K}-2) g(N+1, \bar{K}-1) \varphi(N+1)$ for $N = J - \tfrac{1}{2}$

$\langle J, N-1, \bar{K}+2, S, -\wp | H_{SR} | J, N, \bar{K}, S, \wp \rangle = \tfrac{i}{4} c f(N+1, \bar{K}-2) g(N+1, \bar{K}-1) \varphi(N+1)$ for $N = J - \tfrac{1}{2}$

$\langle J, N-1, \bar{K}-2, S, \wp | H_{SR} | J, N, \bar{K}, S, \wp \rangle = -\tfrac{1}{4} b f(N, \bar{K}-1) g(N, -\bar{K}+1) \varphi(N)$ for $N = J + \tfrac{1}{2}$

$\langle J, N-1, \bar{K}-2, S, -\wp | H_{SR} | J, N, \bar{K}, S, \wp \rangle = \tfrac{i}{4} c f(N, \bar{K}-1) g(N, -\bar{K}+1) \varphi(N)$ for $N = J + \tfrac{1}{2}$

$\langle J, N-1, \bar{K}-2, S, \wp | H_{SR} | J, N, \bar{K}, S, \wp \rangle = -\tfrac{1}{4} b f(N+1, \bar{K}+1) g(N+1, -\bar{K}-1) \varphi(N+1)$ for $N = J - \tfrac{1}{2}$

$\langle J, N-1, \bar{K}-2, S, -\wp | H_{SR} | J, N, \bar{K}, S, \wp \rangle = \tfrac{i}{4} c f(N+1, \bar{K}+1) g(N+1, -\bar{K}-1) \varphi(N+1)$ for $N = J - \tfrac{1}{2}$

## S.8. Line strength and selection rules for transitions between an isolated $\Lambda = 0$ state and nearly degenerate $\Lambda = \pm 1$ states in the orbitally symmetrized case (a) basis set:

Without losing generality, we chose the $\Lambda = 0$ state to be the initial state, and the nearly degenerate $\Lambda = \pm 1$ states to be the final states, so that in the orbitally symmetrized case (a) basis set, the basis functions of the initial and final states are:

$$|\Phi_i\rangle = |\Lambda' = 0\rangle |J'P'M'\rangle |S'\Sigma'\rangle, \tag{1}$$

and

$$|\Phi_f\rangle = \tfrac{1}{\sqrt{2}}|\Lambda = +1\rangle |JPM\rangle |S\Sigma\rangle + \Gamma |\Lambda = -1\rangle |JPM\rangle |S\Sigma\rangle. \tag{2}$$

Quantum numbers of the $\Lambda = 0$ state are primed. Basis functions are written as direct products of the orbital, rotational, and spin basis functions. $M$, the projection of the total angular moment $\boldsymbol{J}$ onto the space-fixed $Z$ axis, is included explicitly.

The line strength for a transition between these two states is (see Eq. (45) in the main text):

$$S(\Gamma', J', P', S', \Sigma'; \Gamma, J, P, S, \Sigma) = 3 \sum_{M,M'} \left| \sum_{q=0,\pm 1} \sum_{i,f} a_i a_f \langle \Phi_i | D_{0q}^{1*} T_q^1(\mu) | \Phi_f \rangle \right|^2$$

$$= \frac{3}{2} \sum_{M,M'} \left| \sum_{q=0,\pm 1} \sum_{i,f} a_i a_f \langle \Lambda' = 0 | \langle J'P'M' | \langle S'\Sigma' | D_{0q}^{1*} T_q^1(\mu) \times (|\Lambda = +1\rangle |JPM\rangle |S\Sigma\rangle + \Gamma |\Lambda = -1\rangle |JPM\rangle |S\Sigma\rangle) \right|^2$$

$$= \frac{3}{2} \sum_{M,M'} \left| \sum_{q=0,\pm 1} \sum_{i,f} a_i a_f \langle S'\Sigma' | S\Sigma \rangle \begin{bmatrix} \langle J'P'M' | D_{0q}^{1*} | JPM \rangle \langle \Lambda' = 0 | T_q^1(\mu) | \Lambda = +1\rangle \\ + \langle J'P'M' | D_{0q}^{1*} | JPM \rangle \Gamma \langle \Lambda' = 0 | \mu_q^1 | \Lambda = -1 \rangle \end{bmatrix} \right|^2 \tag{3}$$

One can then switch all $q$'s to $-q$'s in the second term in the square brackets in Eq. (3), which doesn't change the summation:

$$S(\Gamma', J', P', S', \Sigma'; \Gamma, J, P, S, \Sigma) =$$

$$= \frac{3}{2} \sum_{M,M'} \left| \sum_{q=0,\pm 1} \sum_{i,f} a_i a_f \langle S'\Sigma' | S\Sigma \rangle \begin{bmatrix} \langle J'P'M' | D_{0q}^{1*} | JPM \rangle \langle \Lambda' = 0 | T_q^1(\mu) | \Lambda = +1\rangle \\ + \langle J'P'M' | D_{0-q}^{1*} | JPM \rangle \Gamma \langle \Lambda' = 0 | \mu_{-q}^1 | \Lambda = -1 \rangle \end{bmatrix} \right|^2. \tag{4}$$

Using the symmetry properties of $T_q^1(\mu)$, it can be proven that:[4, 5]

$$\langle \Lambda' = 0 | T_{-q}^1(\mu) | \Lambda = -1 \rangle = (-1)^q \langle \Lambda' = 0 | T_q^1(\mu) | \Lambda = +1 \rangle. \tag{5}$$

We define the electronic transition dipole moment:

$$\mathbf{M}_q = \langle \Lambda' = 0 | T_q^1(\mu) | \Lambda = +1 \rangle. \tag{6}$$

On substitution of Eq. (5) into Eq. (4), and using the definition of $\mathbf{M}_q$ (Eq. 6), one has:

$$S(\Gamma',J',P',S',\Sigma';\Gamma,J,P,S,\Sigma)$$

$$=\frac{3}{2}\sum_{M,M'}\left|\sum_{q=0,\pm 1}\left\{\sum_{i,f}a_i a_f \langle S'\Sigma'|S\Sigma\rangle\left[\begin{array}{l}\langle J'P'M'|D^{1*}_{0q}|JPM\rangle\\+(-1)^q\Gamma\langle J'P'M'|D^{1*}_{0-q}|JPM\rangle\end{array}\right]\right\}\mathbf{M}_q\right|^2. \quad (7)$$

The rotational matrix element in $S(i;f)$ has been evaluated in literature (see, for instance, Eq. (6.123) of Ref. [6][c]):

$$3\sum_{M,M'}\left|\sum_{q=0,\pm 1}\sum_{i,f}a_i a_f \langle J'P'M'|D^{1*}_{0q}|JPM\rangle\right|^2$$

$$=(2J+1)(2J'+1)\left|\sum_{q=0,\pm 1}\sum_{P,P'}a_i a_f (-1)^{J'+P-1}\begin{pmatrix}J & 1 & J'\\ P & q & -P'\end{pmatrix}\right|^2. \quad (8)$$

Therefore, Eq. (7) becomes:

$$S(\Gamma',J',P',S',\Sigma';\Gamma,J,P,S,\Sigma)$$

$$=\frac{1}{2}(2J+1)(2J'+1)\left|\sum_{q=0,\pm 1}\left\{\sum_{i,f,}a_i a_f \langle S'\Sigma'|S\Sigma\rangle(-1)^{J'+P-1}\left[\begin{array}{l}\begin{pmatrix}J & 1 & J'\\ P & q & -P'\end{pmatrix}\\+(-1)^q\Gamma\begin{pmatrix}J & 1 & J'\\ P & -q & -P'\end{pmatrix}\end{array}\right]\right\}\mathbf{M}_q\right|^2. \quad (9)$$

Furthermore, for electric dipole transitions:

$$\langle S'\Sigma'|S\Sigma\rangle=\delta_{S,S'}\delta_{\Sigma,\Sigma'}. \quad (10)$$

Hence, we obtain the line strength formula for transitions between a $\Lambda=0$ state and nearly degenerate $\Lambda=\pm 1$ states in the orbitally symmetrized case (a) basis set:

$$S(\Gamma',J',P',S',\Sigma';\Gamma,J,P,S,\Sigma)$$

$$=\frac{1}{2}\delta_{S,S'}(2J+1)(2J'+1)\left|\sum_{q=0,\pm 1}\left\{\sum_{i,f}a_i a_f \delta_{\Sigma,\Sigma'}(-1)^{J'+P-1}\left[\begin{array}{l}\begin{pmatrix}J & 1 & J'\\ P & q & -P'\end{pmatrix}\\+(-1)^q\Gamma\begin{pmatrix}J & 1 & J'\\ P & -q & -P'\end{pmatrix}\end{array}\right]\right\}\mathbf{M}_q\right|^2. \quad (11)$$

Summation $\sum_{i,f}$ is over all basis functions of both the initial and the final states with the same good quantum numbers ($J/J'$ and $S/S'$), i.e., over all possible values of $\Gamma$, $P/P'$, and $\Sigma/\Sigma'$. Using the line strength formula above, the selection rules for $\Lambda'=0\leftrightarrow\Lambda=\pm 1$ transitions in the orbitally symmetrized case (a) basis set can be derived:

---

[c] Note that a phase factor $(-1)^{J'-1+K''}$ needs to be inserted in front of the 3-$j$ symbol inside the double summation in Eq. (6.123) of Ref. [3]).

(i)      $J' = J, J \pm 1$
(ii)     $P' = P, P \pm 1$ for $q = 0, \pm 1$
(iii)    $S' = S$
(iv)    $\Sigma' = \Sigma$

The second selection rule, which is related to transition types, deserves more discussion. $M_q$'s are matrix elements of the first-rank irreducible tensor of the electronic transition dipole operator in a spherical basis set. Specifically,

$$\mathbf{M}_0 = \langle \Lambda' = 0 | \mu_{z'} | \Lambda = +1 \rangle = \langle \Lambda' = 0 | \mu_{z'} | \Lambda = -1 \rangle$$
$$\mathbf{M}_{\pm 1} = \mp \tfrac{1}{\sqrt{2}} \langle \Lambda' = 0 | \mu_{x'} \pm i\mu_{y'} | \Lambda = +1 \rangle = \pm \tfrac{1}{\sqrt{2}} \langle \Lambda' = 0 | \mu_{x'} \mp i\mu_{y'} | \Lambda = -1 \rangle . \tag{12}$$

Due to the symmetry of orbital wave functions, $\mathbf{M}_0$ vanishes for the $\Lambda' = 0 \leftrightarrow \Lambda = \pm 1$ transitions under investigation. With $q = \pm 1$, the term in the square brackets in Eq. (11) becomes $\begin{pmatrix} J & 1 & J' \\ P & q & -P' \end{pmatrix} - \Gamma \begin{pmatrix} J & 1 & J' \\ P & -q & -P' \end{pmatrix}$, i.e., $\begin{pmatrix} J & 1 & J' \\ P & q & -P' \end{pmatrix} \mp \begin{pmatrix} J & 1 & J' \\ P & -q & -P' \end{pmatrix}$ for $\Gamma = \pm 1$. As a result, for transitions from or to the $\Gamma = \pm 1$ basis functions the line strength $S$ is proportional to $(\mathbf{M}_{+1} \mp \mathbf{M}_{-1})$ (see Eq. (11)). It has been proven that $(\mathbf{M}_{+1} - \mathbf{M}_{-1})$ and $(\mathbf{M}_{+1} + \mathbf{M}_{-1})$ are equivalent to $\langle \Lambda' = 0 | \mu_{x'} | \Gamma = +1 \rangle$ and $\langle \Lambda' = 0 | \mu_{y'} | \Gamma = -1 \rangle$, respectively, where $\mu_{x'}$ and $\mu_{y'}$ are electronic transition dipole moments along the $x'$ and the $y'$ axes, respectively. (See Section S.1.3 in the Supporting Information of Ref. [7] for proof.) Therefore it is evident that for transitions from or to the $\Gamma = +1$ basis functions, i.e., the $A'$ state, the electronic transition dipole is along the $x'$ axis, whereas it is along the $y'$ axis for transitions from or to the $\Gamma = -1$ basis functions, i.e., the $A''$ state. The transition dipole moment can be readily converted from the IAS to PAS through unitary transformation of coordinate systems.

**S.9. Line strength and selection rules for transitions between an isolated $\Lambda = 0$ state and nearly degenerate $\Lambda = \pm 1$ states in the orbitally symmetrized case (b) basis set:**

In the orbitally symmetrized case (a) basis set:

$$|\Phi_i\rangle = |J',N',K',S'\rangle, \tag{13}$$

and

$$|\Phi_f\rangle = \tfrac{1}{\sqrt{2}}\big[|\Lambda=+1\rangle|J,N,K,S\rangle + \Gamma|\Lambda=-1\rangle|J,N,K,S\rangle\big]. \tag{14}$$

Line strength of transitions between two *isolated* states in the case (b) spin-rotational basis set has been derived previously.[8] In the orbitally symmetrized basis set, the summation should be over not only $N/N'$ but also $\Gamma$. Similar to derivation in Section S.8, the line strength for a transition between the two states in the orbitally symmetrized case (b) basis set is found to be:

$$S(\Gamma',J',N',K',S';\Gamma,J,N,K,S)$$
$$= \frac{1}{2}\delta_{S,S'}(2J+1)(2J'+1)\left|\sum_{\Gamma}\sum_{N,N'}(-1)^{J'+N+S+1}\begin{pmatrix}N & J & S \\ J' & N' & 1\end{pmatrix}\tilde{S}^{1/2}(J',N',K',S';J,N,K,S)\right|^2, \tag{15}$$

where

$$\tilde{S}^{1/2}(J',N',K',S';J,N,K,S)$$
$$= (2N+1)^{1/2}(2N'+1)^{1/2}\sum_{q=0,\pm 1}\left\{\sum_{K,K'}a_i a_f (-1)^{N-K'-1}\left[\begin{pmatrix}N & 1 & N' \\ K & q & -K'\end{pmatrix} + (-1)^q\Gamma\begin{pmatrix}N & 1 & N' \\ K & -q & -K'\end{pmatrix}\right]\right\}\mathbf{M}_q. \tag{16}$$

The selection rules are:

(i) $J' = J, J \pm 1$

(ii) $N' = N, N \pm 1$

(iii) $K' = K, K \pm 1$ for $q = 0, \pm 1$

Arguments on transition types in Section S.8 apply to the orbitally symmetrized case (b) basis set as well.

**S.10. Intensity formula and selection rules for transitions between an isolated $\Lambda = 0$ state and nearly degenerate $\Lambda = \pm 1$ states in the fully symmetrized case (a) basis set:**

In the fully symmetrized case (a) basis set:

$$|\Phi_i\rangle = |J', \bar{P}', S', \bar{\Sigma}', \wp'\rangle$$
$$= \tfrac{1}{\sqrt{2}} |\Lambda' = 0\rangle \left[ |J'P'M'\rangle |S'\Sigma'\rangle + s'\wp' |J'-P'M'\rangle |S'-\Sigma'\rangle \right], \qquad (17)$$

and

$$|\Phi_f\rangle = |J, \bar{P}, S, \bar{\Sigma}, \wp\rangle$$
$$= \tfrac{1}{\sqrt{2}} \left[ |\Lambda = +1\rangle |JPM\rangle |S\Sigma\rangle + s\wp |\Lambda = -1\rangle |J-PM\rangle |S-\Sigma\rangle \right], \qquad (18)$$

where $s = (-1)^{J-P+S-\Sigma}$ and $s' = (-1)^{J'-P'+S'-\Sigma'}$. Note that for the $\Lambda = 0$ state, energy levels with $\wp' = \pm 1$ are degenerate (see main text).

If $\bar{\Sigma}' = \bar{\Sigma}$, the line strength for a transition between the two states is:

$$S(J', \bar{P}', S', \bar{\Sigma}' = \bar{\Sigma}, \wp'; J, \bar{P}, S, \bar{\Sigma}, \wp) = 3 \sum_{M,M'} \left| \sum_{q=0,\pm 1} \sum_{i,f} a_i a_f \langle \Phi_i | D^{1*}_{0q} T^1_q(\mu) | \Phi_f \rangle \right|^2$$

$$= \frac{3}{4} \sum_{M,M'} \left| \sum_{q=0,\pm 1} \sum_{i,f} a_i a_f \begin{bmatrix} \langle \Lambda' = 0 | \langle J'P'M' | \langle S'\Sigma' | D^{1*}_{0q} T^1_q(\mu) | \Lambda = +1\rangle |JPM\rangle |S\Sigma\rangle \\ + ss'\wp\wp' \langle \Lambda' = 0 | \langle J'-P'M' | \langle S'-\Sigma' | D^{1*}_{0q} T^1_q(\mu) | \Lambda = -1\rangle |J-PM\rangle |S-\Sigma\rangle \end{bmatrix} \right|^2$$

$$= \frac{1}{4} \delta_{S,S'} (2J+1)(2J'+1) \left| \sum_{q=0,\pm 1} \sum_{\wp,\wp'} \sum_{P,P'} a_i a_f \begin{bmatrix} (-1)^{J'+P-1} \begin{pmatrix} J & 1 & J' \\ P & q & -P' \end{pmatrix} \langle \Lambda' = 0 | \mu^1_q | \Lambda = +1 \rangle \\ + ss'\wp\wp'(-1)^{J'-P-1} \begin{pmatrix} J & 1 & J' \\ -P & q & P' \end{pmatrix} \langle \Lambda' = 0 | \mu^1_q | \Lambda = -1 \rangle \end{bmatrix} \right|^2$$

(19)

Relations in Eqs. (8) and (10) are applied in the derivation above. Eq. (19) can be further simplified by switching $q$'s to $-q$'s in the second term in the square brackets, which doesn't change the summation, and then using the symmetry property of the 3-$j$ symbols:

$$\begin{pmatrix} J & 1 & J' \\ -P & q & P' \end{pmatrix} = (-1)^{J+J'+1} \begin{pmatrix} J & 1 & J' \\ P & -q & -P' \end{pmatrix}. \qquad (20)$$

One therefore obtains:

$$S(J',\bar{P}',S',\bar{\Sigma}'=\bar{\Sigma},\wp';J,\bar{P},S,\bar{\Sigma},\wp)$$

$$=\frac{1}{4}\delta_{S,S'}(2J+1)(2J'+1)\left|\sum_{q=0,\pm 1}\sum_{\wp,\wp'}\sum_{P,P'}a_i a_f \begin{bmatrix}(-1)^{J'+P-1}\langle\Lambda'=0|\mu_q^1|\Lambda=+1\rangle+ss'\wp\wp'\\ \times(-1)^{J'-P-1}(-1)^{J+J'+1}\langle\Lambda'=0|\mu_{-q}^1|\Lambda=-1\rangle\end{bmatrix}\begin{pmatrix}J & 1 & J'\\ P & q & -P'\end{pmatrix}\right|^2$$

$$=\frac{1}{4}\delta_{S,S'}(2J+1)(2J'+1)\left|\sum_{q=0,\pm 1}\left\{\sum_{\wp,\wp'}\sum_{P,P'}a_i a_f (-1)^{J'+P-1}\begin{bmatrix}1+ss'\wp\wp'\\ \times(-1)^{-2P}(-1)^{J+J'+1}(-1)^q\end{bmatrix}\begin{pmatrix}J & 1 & J'\\ P & q & -P'\end{pmatrix}\right\}\mathbf{M}_q\right|^2$$

(21)

The second step in the derivation above uses the relation in Eq. (5) and the definition of $\mathbf{M}_q$ (Eq. 6). Several other relations can be used to further simplifies Eq. (21). First, the phase factor

$$ss'(-1)^{-2P}(-1)^{J+J'+1}(-1)^q = (-1)^{2J+2J'-3P-P'+S+S'-\Sigma-\Sigma'+1+q}. \tag{22}$$

In the present work, $S=S'=1/2$ so that $J$, $J'$, $P$, $P'$, $\Sigma$, and $\Sigma'$ are all half-integers. As a result $2J+2J'$ is even, while $2P$, $S+S'$ and $\Sigma+\Sigma'=2\Sigma$ are odd. Thus:

$$ss'(-1)^{-2P}(-1)^{J+J'+1}(-1)^q = (-1)^{-P-P'+q}. \tag{23}$$

In order for the 3-$j$ symbol in Eq. (21) to be nonzero, the condition $P'=P+q$ has to be satisfied. The phase factor is therefore equal to $(-1)^{2P}=-1$, and Eq. (21) reduces to:

$$S(J',\bar{P}',S',\bar{\Sigma}'=\bar{\Sigma},\wp';J,\bar{P},S,\bar{\Sigma},\wp)$$

$$=\frac{1}{4}\delta_{S,S'}(2J+1)(2J'+1)\left|\sum_{q=0,\pm 1}\left\{\sum_{\wp,\wp'}\sum_{P,P'}a_i a_f (-1)^{J'+\bar{P}-1}[1-\wp\wp']\begin{pmatrix}J & 1 & J'\\ -\bar{P} & q & \bar{P}'\end{pmatrix}\right\}\mathbf{M}_q\right|^2. \tag{24}$$

It is obvious that $1-\wp\wp'=2\delta_{\wp,-\wp'}$, which leads to the line strength formula for the case of $\bar{\Sigma}'=\bar{\Sigma}$:

$$S(J',\bar{P}',S',\bar{\Sigma}'=\bar{\Sigma},\wp';J,\bar{P},S,\bar{\Sigma},\wp)$$

$$=\frac{1}{2}\delta_{S,S'}(2J+1)(2J'+1)\left|\sum_{q=0,\pm 1}\left\{\sum_{\wp,\wp'}\sum_{\bar{P},\bar{P}'}a_i a_f (-1)^{J'+\bar{P}-1}\delta_{\wp,-\wp'}\begin{pmatrix}J & 1 & J'\\ -\bar{P} & q & \bar{P}'\end{pmatrix}\right\}\mathbf{M}_q\right|^2. \tag{25}$$

If $\bar{\Sigma}'=-\bar{\Sigma}$,

$$S(J',\bar{P}',S',\bar{\Sigma}'=-\bar{\Sigma},\wp';J,\bar{P},S,\bar{\Sigma},\wp)$$

$$=\frac{3}{4}\sum_{M,M'}\left|\sum_{q=0,\pm 1}\sum_{i,f}a_i a_f \begin{bmatrix}s\wp\langle\Lambda'=0|\langle J'P'M'|\langle S'\Sigma'|D_{0q}^{1*}T_q^1(\mu)|\Lambda=-1\rangle|J-PM\rangle|S-\Sigma\rangle\\ +s'\wp'\langle\Lambda'=0|\langle J'-P'M'|\langle S'-\Sigma'|D_{0q}^{1*}T_q^1(\mu)|\Lambda=+1\rangle|JPM\rangle|S\Sigma\rangle\end{bmatrix}\right|^2$$

$$=\frac{3}{4}\sum_{M,M'}\left|\sum_{q=0,\pm 1}\sum_{i,f}a_i a_f s\wp\begin{bmatrix}\langle\Lambda'=0|\langle J'P'M'|\langle S'\Sigma'|D_{0q}^{1*}T_q^1(\mu)|\Lambda=-1\rangle|J-PM\rangle|S-\Sigma\rangle\\ +ss'\wp\wp'\langle\Lambda'=0|\langle J'-P'M'|\langle S'-\Sigma'|D_{0q}^{1*}T_q^1(\mu)|\Lambda=+1\rangle|JPM\rangle|S\Sigma\rangle\end{bmatrix}\right|^2$$

(26)

Applying the relations employed in the $\overline{\Sigma}' = \overline{\Sigma}$ case, one obtains the line strength formula for the case of $\overline{\Sigma}' = -\overline{\Sigma}$:

$$S(J',\overline{P}',S',\overline{\Sigma}'=-\overline{\Sigma},\wp';J,\overline{P},S,\overline{\Sigma},\wp)$$
$$=\frac{1}{2}\delta_{S,S'}(2J+1)(2J'+1)\left|\sum_{q=0,\pm 1}\left\{\sum_{\wp,\wp'}\sum_{P,P'}a_i a_f s'\wp'(-1)^{J'+\overline{P}-1}\delta_{\wp,-\wp'}\begin{pmatrix}J & 1 & J'\\ \overline{P} & q & \overline{P}'\end{pmatrix}\right\}\mathbf{M}_q\right|^2. \qquad (27)$$

Overall line strength can be obtained by combining the $\overline{\Sigma}' = \overline{\Sigma}$ and $\overline{\Sigma}' = -\overline{\Sigma}$ cases and written as:

$$S(J',\overline{P}',S',\overline{\Sigma}',\wp';J,\overline{P},S,\overline{\Sigma},\wp)$$
$$=\frac{1}{2}\delta_{S,S'}(2J+1)(2J'+1)\left|\sum_{q=0,\pm 1}\left[\sum_{\wp,\wp'}\sum_{\overline{P},\overline{P}'}\sum_{\overline{\Sigma},\overline{\Sigma}'}a_i a_f \tilde{F}^{1/2}(J',\overline{P}',S',\overline{\Sigma}',\wp';J,\overline{P},S,\overline{\Sigma},\wp)\right]\mathbf{M}_q\right|^2, \qquad (28)$$

where

$$\tilde{F}^{1/2}(J',\overline{P}',S',\overline{\Sigma}',\wp';J,\overline{P},S,\overline{\Sigma},\wp) = \begin{cases} (-1)^{J'+\overline{P}-1}\delta_{\wp,-\wp'}\begin{pmatrix}J & 1 & J'\\ -\overline{P} & q & \overline{P}'\end{pmatrix} & \text{if } \overline{\Sigma}' = \overline{\Sigma} \\ s'\wp'(-1)^{J'+\overline{P}-1}\delta_{\wp,-\wp'}\begin{pmatrix}J & 1 & J'\\ \overline{P} & q & \overline{P}'\end{pmatrix} & \text{if } \overline{\Sigma}' = -\overline{\Sigma} \end{cases}. \qquad (29)$$

It is worth noting that $s'\wp' = \Gamma'$. In the present work, the initial state (with primed quantum numbers) is a $\Lambda = 0$ state. Therefore $\Gamma'$ is nominal and can be set to 1. The line strength formulae derived above can be used directly in the case of $\Lambda' = \pm 1 \leftrightarrow \Lambda = \pm 1$ transitions.

Selection rules for $\Lambda' = 0 \leftrightarrow \Lambda = \pm 1$ transitions in the fully symmetrized case (a) basis set can be determined from Eqs. (25) and (27):

(i)      $J' = J, J \pm 1$

(ii)     $\overline{P}' = \overline{P}, \overline{P} \pm 1$ for $q = 0, \mp 1$, if $\overline{\Sigma}' = \overline{\Sigma}$, and

         $\overline{P}' = -\overline{P}, -\overline{P} \pm 1$ for $q = 0, \mp 1$, if $\overline{\Sigma}' = -\overline{\Sigma}$,

(iii)    $S' = S$

(iv)    $\wp' = -\wp$

**S.11. Intensity formula and selection rules for transitions between an isolated $\Lambda = 0$ state and nearly degenerate $\Lambda = \pm 1$ states in the fully symmetrized case (b) basis set:**

In the fully symmetrized case (a) basis set:

$$|\Phi_i\rangle = |J',N',\bar{K}',S',\wp'\rangle$$
$$= \tfrac{1}{\sqrt{2}}|\Lambda'=0\rangle\left[|J',N',K',S'\rangle + s'\wp'|J',N',-K',S'\rangle\right], \qquad (30)$$

and

$$|\Phi_f\rangle = |J,N,\bar{K},S,\wp\rangle$$
$$= \tfrac{1}{\sqrt{2}}\left[|\Lambda=+1\rangle|J,N,K,S\rangle + s\wp|\Lambda=-1\rangle|J,N,-K,S\rangle\right], \qquad (31)$$

where $s = (-1)^{N-K}$ and $s' = (-1)^{N'-K'}$.

The line strength for a transition between the two states is a sum of four terms:

$$S(J',N',\bar{K}',S',\wp';J,N,\bar{K},S,\wp) =$$
$$= \frac{1}{4}\delta_{S,S'}(2J+1)(2J'+1)\left|\sum_{\wp\wp'}\sum_{N,N'}(-1)^{J'+N+S+1}\begin{pmatrix} N & J & S \\ J' & N' & 1 \end{pmatrix}\sum_{n=1}^{4}\tilde{S}_n^{1/2}(J',N',K',S';J,N,K,S)\right|^2, \qquad (32)$$

where

$$\tilde{S}_1^{1/2}(J',N',\bar{K}',S';J,N,\bar{K},S)$$
$$= (2N+1)^{1/2}(2N'+1)^{1/2}\sum_{q=0,\pm1}\sum_{K,K'}a_i a_f (-1)^{N-\bar{K}'-1}\begin{pmatrix} N & 1 & N' \\ \bar{K} & q & -\bar{K}' \end{pmatrix}\langle\Lambda'=0|T_q^1(\mu)|\Lambda=+1\rangle$$

$$\tilde{S}_2^{1/2}(J',N',\bar{K}',S';J,N,\bar{K},S)$$
$$= (2N+1)^{1/2}(2N'+1)^{1/2}\sum_{q=0,\pm1}\sum_{K,K'}a_i a_f s'\wp'(-1)^{N+\bar{K}'-1}\begin{pmatrix} N & 1 & N' \\ \bar{K} & q & \bar{K}' \end{pmatrix}\langle\Lambda'=0|T_q^1(\mu)|\Lambda=+1\rangle$$

$$\tilde{S}_3^{1/2}(J',N',\bar{K}',S';J,N,\bar{K},S) \qquad (33)$$
$$= (2N+1)^{1/2}(2N'+1)^{1/2}\sum_{q=0,\pm1}\sum_{q=0,\pm1}a_i a_f s\wp(-1)^{N-\bar{K}'-1}\begin{pmatrix} N & 1 & N' \\ -\bar{K} & q & -\bar{K}' \end{pmatrix}\langle\Lambda'=0|T_q^1(\mu)|\Lambda=-1\rangle$$

$$\tilde{S}_4^{1/2}(J',N',\bar{K}',S';J,N,\bar{K},S)$$
$$= (2N+1)^{1/2}(2N'+1)^{1/2}\sum_{q=0,\pm1}\sum_{q=0,\pm1}a_i a_f ss'\wp\wp'(-1)^{N+\bar{K}'-1}\begin{pmatrix} N & 1 & N' \\ -\bar{K} & q & \bar{K}' \end{pmatrix}\langle\Lambda'=0|T_q^1(\mu)|\Lambda=-1\rangle$$

Applying the same techniques used in Section S.10, it can be proven that:

$$\sum_{n=1}^{4} \tilde{S}_n^{1/2}(J',N',\bar{K}',S';J,N,\bar{K},S)$$

$$= 2\delta_{\wp,-\wp'}[1+s'\wp'](2N+1)^{1/2}(2N'+1)^{1/2} \sum_{q=0,\pm 1} \left\{ \sum_{K,K'} a_i a_f (-1)^{N-K'-1} \begin{pmatrix} N & 1 & N' \\ \bar{K} & q & -\bar{K}' \end{pmatrix} \right\} \mathbf{M}_q . \quad (34)$$

One eventually obtains the line strength formula for a transition between the two states in the fully symmetrized case (b) basis set:

$$S(J',N',\bar{K}',S',\wp';J,N,\bar{K},S,\wp) =$$

$$= \frac{1}{4}\delta_{S,S'}(2J+1)(2J'+1) \left| \sum_{\wp\wp'} \sum_{N,N'} (-1)^{J'+N+S+1} \begin{pmatrix} N & J & S \\ J' & N' & 1 \end{pmatrix} \tilde{S}^{1/2}(J',N',K',S';J,N,K,S) \right|^2, \quad (35)$$

where

$$\tilde{S}^{1/2}(J',N',\bar{K}',S';J,N,\bar{K},S)$$

$$= 2\delta_{\wp,-\wp'}[1+s'\wp'](2N+1)^{1/2}(2N'+1)^{1/2} \sum_{q=0,\pm 1} \left\{ \sum_{K,K'} a_i a_f (-1)^{N-K'-1} \begin{pmatrix} N & 1 & N' \\ \bar{K} & q & -\bar{K}' \end{pmatrix} \right\} \mathbf{M}_q . \quad (36)$$

Selection rules with this basis sets are:

(i)      $J' = J, J \pm 1$

(ii)     $N' = N, N \pm 1$

(iii)    $\bar{K}' = \bar{K}, \bar{K} \pm 1$ for $q = 0, \pm 1$

(iv)    $\wp' = -\wp$

**References:**

1. J. H. Van Vleck, Rev. Mod. Phys. **23**, 213 (1951).
2. W. T. Raynes, J. Chem. Phys. **41**, 3020 (1964).
3. J. M. Brown and T. J. Sears, J. Mol. Spectrosc. **75**, 111 (1979).
4. J. Liu, Ph.D. Thesis, The Ohio State University, 2007.
5. I. J. Kalinovski, Ph.D. Thesis, University of California, Berkeley, 2001.
6. R. N. Zare, *Angular Momentum*. (Wiley, New York, 1988)
7. J. Liu and T. A. Miller, J. Phys. Chem. A **118**, 11871 (2014).
8. S. Gopalakrishnan, C. C. Carter, L. Zu, V. Stakhursky, G. Tarczay and T. A. Miller, J. Chem. Phys. **118**, 4954 (2003).